\documentclass{article}

\PassOptionsToPackage{numbers, compress}{natbib}

\usepackage[final]{neurips_2020}

\usepackage[utf8]{inputenc} 
\usepackage[T1]{fontenc}    
\usepackage{hyperref}       
\usepackage{url}            
\usepackage{booktabs}       
\usepackage{amsfonts}       
\usepackage{nicefrac}       
\usepackage{microtype}      
\usepackage{graphicx}
\usepackage{amsmath,amsfonts,amssymb,amsthm}
\usepackage{array}
\usepackage{subcaption}
\usepackage{bbm}
\usepackage{bm}
\usepackage{multicol}
\usepackage{lipsum}
\usepackage{wrapfig}
\usepackage[ruled]{algorithm2e}
\usepackage{arydshln}
\usepackage{multirow}

\newcommand{\inv}{^{\raisebox{.2ex}{$\scriptscriptstyle-1$}}}

\newtheorem{claim}{Claim}

\newtheorem{theorem}{Theorem}
\newtheorem{corollary}{Corollary}

\newcommand{\Ebb}{\mathbb{E}}
\newcommand{\Rbb}{\mathbb{R}}
\newcommand{\Ncal}{\mathcal{N}}
\newcommand{\Dcal}{\mathcal{D}}
\newcommand{\Xcal}{\mathcal{X}}
\newcommand{\Gcal}{\mathcal{G}}
\newcommand{\Fcal}{\mathcal{F}}
\newcommand{\Jcal}{\mathcal{J}}

\newcommand{\tbf}{\mathbf{t}}

\newcommand{\xbf}{\mathbf{x}}
\newcommand{\Xbf}{\mathbf{X}}
\newcommand{\zbf}{\mathbf{z}}
\newcommand{\Zbf}{\mathbf{Z}}

\title{Adversarial Counterfactual Learning and Evaluation for Recommender System}

\author{Da Xu, Chuanwei Ruan \thanks{Both authors contribute equally to this work.} \\ Walmart Labs, Sunnyvale, CA 94086 \\ \{Da.Xu, Chuanwei.Ruan\}@walmartlabs.com 
\And 
Evren Korpeoglu, Sushant Kumar, Kannan Achan \\
Walmart Labs, Sunnyvale, CA 94086 \\
\{EKorpeoglu, SKumar4, KAchan\}@walmartlabs.com}

\begin{document}

\maketitle

\begin{abstract}
The feedback data of recommender systems are often subject to what was exposed to the users; however, most learning and evaluation methods do not account for the underlying exposure mechanism. We first show in theory that applying supervised learning to detect user preferences may end up with inconsistent results in the absence of exposure information. The counterfactual propensity-weighting approach from causal inference can account for the exposure mechanism; nevertheless, the partial-observation nature of the feedback data can cause identifiability issues. We propose a principled solution by introducing a minimax empirical risk formulation. We show that the relaxation of the dual problem can be converted to an adversarial game between two recommendation models, where the opponent of the candidate model characterizes the underlying exposure mechanism. We provide learning bounds and conduct extensive simulation studies to illustrate and justify the proposed approach over a broad range of recommendation settings, which shed insights on the various benefits of the proposed approach.
\end{abstract}

\section{Introduction}
\label{sec:intro}
\begin{figure}
    \centering
    \includegraphics[scale=0.14]{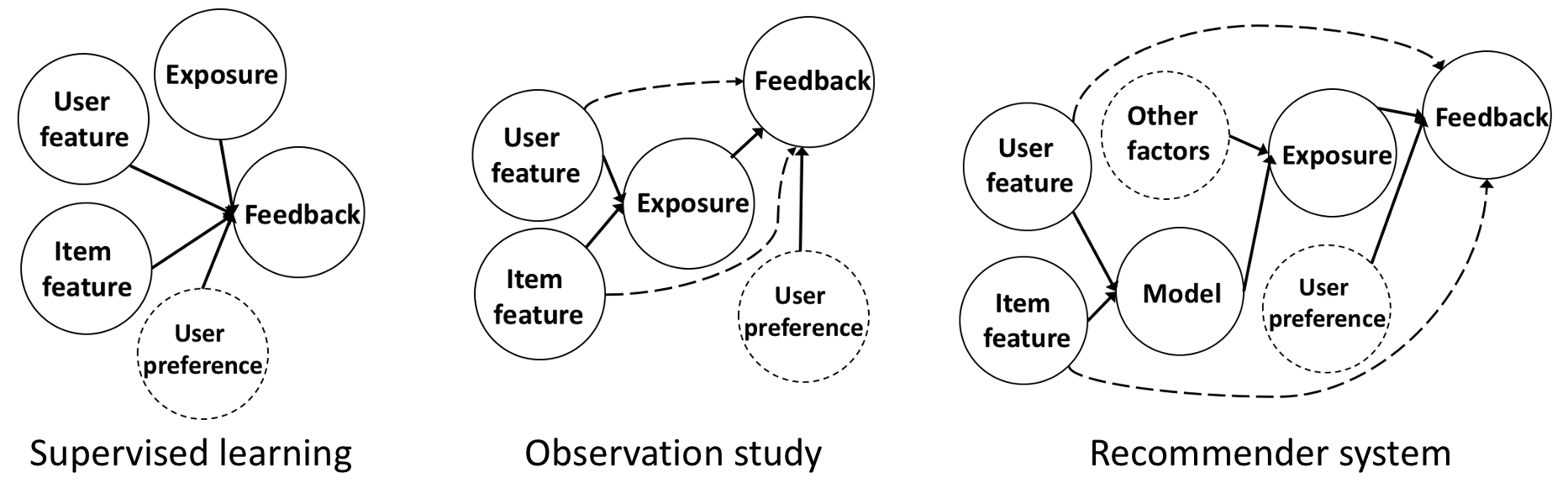}
    \caption{The graphical model representation for the causal inference view under different settings.}
    \label{fig:pgm}
\end{figure}

In the offline learning and evaluation of recommender systems, the dependency of feedback data on the underlying exposure mechanism is often overlooked. When the users express their preferences on the products explicitly (such as providing ratings) or implicitly (such as clicking), the feedback are conditioned on the products to which they are exposed. In most cases, the previous exposures are decided by some underlying mechanism such as the history recommender system. The dependency causes two dilemmas for machine learning in recommender systems, and solutions have yet been found satisfactorily.
Firstly, the majority of supervised learning models only handle the dependency between label (user feedback) and features, yet in the actual feedback data, the exposure mechanism can alter the dependency pathways (Figure \ref{fig:pgm}). In Section \ref{sec:prelim}, we show from a theoretical perspective that directly applying supervised learning on feedback data can result in inconsistent detection of the user preferences. Secondly, an unbiased model evaluation should have the product exposure determined by the candidate recommendation model, which is almost never satisfied using the feedback data only. The second dilemma also reveals a major gap between evaluating models by online experiments and using history data, since the offline evaluations are more likely to bias toward the history exposure mechanism as it decided to what products the users might express their preferences.
The disagreement between the online and offline evaluations may partly explain the controversial observations made in several recent papers, where deep recommendation models are overwhelmed by classical collaborative filtering approaches in offline evaluations \cite{dacrema2019we,rendle2020neural}, despite their many successful deployments in the real-world applications \cite{cheng2016wide,covington2016deep,ying2018graph,zhou2018deep,zheng2018drn,zhang2019deep}.

To settle the above dilemmas for recommender systems, we refer to the idea of \emph{counterfactual} modelling from the observational studies and causal inference literature \cite{morgan2015counterfactuals,pearl2009causal,rosenbaum2010design} to redesign the learning and evaluation methods. Briefly put, the counterfactual modelling answers questions related to "what if", e.g. what is the feedback data if the candidate model were deployed. Our key purpose of introducing the counterfactual methods is to take account of the dependency between the feedback data and exposure. Relevant proposals have been made in several recent papers \cite{schnabel2016recommendations,liang2016causal,joachims2017unbiased,agarwal2018counterfactual,liang2016modeling,yang2018unbiased,hernandez2014probabilistic}; however, most of them rely on excessive data or model assumptions (such as the missing-data model we describe in Section \ref{sec:prelim}) that may not be satisfied in practice. Many of the assumptions are essentially unavoidable due to a fundamental discrepancy between the recommender system and observational studies. In observational studies, the exposure (treatment) status are fully observed, and the exposure mechanism is completely decided by the covariates (features) \cite{rosenbaum1983central,austin2011introduction}.
For recommender systems, the exposure is only partially captured by the feedback data. The complete exposure status can only be retrieved from the system's backend log, whose access is highly restricted, and rarely exists for the public datasets.
Also, the exposure mechanism can depend on intractable randomness, e.g. burst events, special offers, interference with other modules such as the advertisement, as well as the relevant features that are not attainable from feedback data. In Figure \ref{fig:pgm}, we show the causal diagrams for the three different views of recommender system.
A direct consequence of the above differences is that the exposure mechanism is not \emph{identifiable} from feedback data, i.e. we can modify the conditional distribution characterized by the exposure mechanism without disturbing the observation distribution. Therefore, the existing methods have to make problem-specific or unjustifiable assumptions in order to bypass or simply ignore the identifiability issue. 

Our solution is to acknowledge the uncertainty brought by the identifiability issue and treat it as an adversarial component. We propose a minimax setting where the candidate model is optimized over the worst-case exposure mechanism. By applying duality arguments and relaxations, we show that the minimax problem can be converted to an \emph{adversarial game} between two recommendation models. Our approach is entirely novel and principled. We conclude the contributions as follow.
\begin{itemize}
    \item We provide the first theoretical analysis to show an inconsistent issue of supervised learning on recommender systems, caused by the unknown exposure mechanism.
    \item We propose a minimax setting for counterfactual recommendation and convert it to a tractable two-model adversarial game. We prove the generalization bounds for the proposed adversarial learning, and provide analysis for the minimax optimization.
    \item We carry out extensive simulation and real data experiments to demonstrate our performance, and deploy online experiments to fully illustrate the benefits of the proposed approach.
\end{itemize}



\section{Preliminaries}
\label{sec:prelim}
We use bold-faced letters to denote vectors and matrices, upper-case letters to denote random variables and the corresponding lower-case letters to denote observations. Distributions are denoted by $P$ and $Q$. Let $\xbf_u$ be the user feature vector for user $u\in \{1,\ldots,n\}$, $\zbf_i$ be the item feature vector for item $i\in \{1,\ldots,m\}$, $O_{u,i} \in \{0,1\}$ be the exposure status, $Y_{u,i}$ be the feedback and $\Dcal$ be the collected user-item pairs where non-positive interactions may come from negative sampling. The feature vectors can be one-hot encoding or embedding, so our approach is fully compatible with deep learning models that leverage representation learning and are trained under negative sampling. Recommendation models are denoted by such as $f_{\theta}$ and $g_{\psi}$. They take $\xbf_u$, $\zbf_i$ (and the exposure $o_{u,i}$ if available) as input. We use the shorthand $f_{\theta}(u,i)$ to denote the output score, and the loss with respect to the $y_{u,i}$ is given by $\delta(y_{u,i}, f_{\theta}(u,i))$. 
Our notations also apply to the sequential recommendation by encoding the previously-interacted items to the user feature vector $\xbf$. 

We use $p_{g}(O_{u,i}|\xbf_u,\zbf_i)$ to denote the exposure mechanism that depends on the underlying model $g$. Also,  $p(Y_{u,i}|O_{u,i},\xbf_{u},\zbf_{i})$ gives the user response, which is independent from the exposure mechanism whenever $O_{u,i}$ is observed. We point out that the stochasticity in the exposure can also be induced by the exogenous factors (unobserved confounders) who bring extra random perturbations. We do not explicitly differentiate the explcit and implicit feedback setting unless specified.



\textbf{Supervised learning for feedback data.} 

Let $Y_{u,i}\in\{-1,1\}$ be the implicit feedback. Set aside the exposure for a moment, the goal of supervised learning is to determine the optimal recommendation function that minimizes the surrogate loss:
$\ell_{\phi}(f_{\theta}) = \frac{1}{|\Dcal|}\sum_{(u,i)\in \Dcal}\big[\phi(Y_{u,i}\cdot f_{\theta}(u,i)) \big]$, where $\phi$ induces the widely-adopted margin-based loss. Now we take account of the (unobserved) exposure status by first letting:
\[
p^{(1)}(o) = p(Y_{u,i}=1, O_{u,i}=o, \xbf_u, \zbf_i), \, p^{(-1)}(o) = p(Y_{u,i}=-1, O_{u,i}=o, \xbf_u, \zbf_i),\, o\in\{0,1\},
\]
to denote the joint distribution for positive and negative feedback under either exposure status. The surrogate loss, which now depends on $p^{(1)}$ and $p^{(-1)}$ since we include the exposure, is denoted by $L_{\phi}\big(f_{\theta}, \{p^{(1)}, p^{(-1)}\}\big)$.
In the following claim, we show that if we fix the exposure mechanism and optimize $f_{\theta}$, the optimal loss and the corresponding $f_{\theta}^*$ depend only on $p^{(1)}$ and $p^{(-1)}$.
\begin{claim}
\label{claim:supervised}
When the exposure mechanism $p(O_{u,i}|\Xbf_u, \Zbf_i)$ is \textbf{given and fixed}, the optimial loss is:
\begin{equation}
\inf_{f_{\theta}}L_{\phi}\big(f_{\theta}, \{p^{(1)}, p^{(-1)}\}\big) = -D_{c}(P^{(1)} || P^{(-1)}),
\end{equation}
where $P^{(1)}$ and $P^{(-1)}$ are the corresponding distributions for $p^{(1)}$ and $p^{(-1)}$, and $D_{c}(P^{(1)} || P^{(-1)})=\int c\big(\frac{p^{(1)}}{p^{(-1)}} \big)dP^{(-1)}$ is the f-divergence induced by the convex, lower-semicontinuous function $c$. Also, the optimal $f_{\theta}^*$ that achieves the infimum is given by $\alpha_{\phi}^*\big(\frac{p^{(1)}}{p^{(-1)}} \big)$ for some function $\alpha_{\phi}^*$ that depends on $\phi$.
\end{claim}
We defer the proof to Appendix \ref{append:claim1}. Notice that the joint distribution can be factorized into: $p(Y_{u,i}, o_{u,i}, \xbf_u, \xbf_i) \propto p(Y_{u,i} | o_{u,i}, \xbf_u, \zbf_i) \cdot p_{g}(o_{u,i}|\xbf_u, \zbf_i)$, so Claim \ref{claim:supervised} implies that:
\[f_{\theta}^*(\xbf_u, \zbf_i; o_{u,i}) = \alpha_{\phi}^* \Big(p(Y_{u,i}=1|o_{u,i}, \xbf_u, \zbf_i)\, \big/ \, p(Y_{u,i}=-1|o_{u,i}, \xbf_u, \zbf_i) \Big).\]
We conclude that: \textbf{1.} when the exposure mechanism is given, the optimal loss $-D_{c}(P^{(1)} || P^{(-1)})$ is a function of both the user preference and the exposure mechanism; \textbf{2.} the optimal model $f_{\theta}^*$ depends only on the user preference, since $f_{\theta}^*$ is a function of $p(Y|o,\xbf,\zbf)$ which does not depend on the exposure mechanism (mentioned at the beginning of this section).
Both conclusions are practically reasonable, as the optimal recommendation model should only detect user preference regardless of the exposure mechanisms. The optimal loss, on the other hand, depends on the joint distribution where the underlying exposure mechanism plays a part. 

However, when $p(O_{u,i}|\Xbf_u, \Zbf_i)$ is unknown, the conclusions from Claim \ref{claim:supervised} no longer hold and the optimal $f_{\theta}^*$ will depend on the exposure mechanism. As a consequence, if the same feedback data were collected under different exposure mechanisms, the recommendation model may find the user preference differently. The inconsistency is caused by not accounting for the unknown exposure mechanism from the supervised learing. We mention that another line of research studies the user preference and exposure in an interactive online fashion using such as the contextual bandit and reinforcement learning \cite{li2010contextual,zheng2018drn}. The discussions of which are beyond the scope of this paper.

\textbf{The propensity-weighting approach}.

In causal inference, the probability of exposure given the observed features (covariates) is referred to as the propensity score \cite{rosenbaum1983central}. The propensity-weighting approach uses weights based on the propensity score to create a synthetic sample in which the distribution of observed features is independent of exposure \cite{hirano2001estimation,austin2011introduction}. It especially appeals to us because we want the feedback data to be made independent of the exposure mechanism. The propensity-weighted loss is constructed via: $\frac{1}{|\Dcal|} \sum_{(u,i)\in\Dcal} \phi\big(y_{u,i}\cdot f_{\theta}(\xbf_u, \zbf_i) \big) \, \big/ \, p(O_{u,i} = 1 | \xbf_u, \zbf_i)$,
and by taking the expectation with respect to exposure (whose distribution is denoted by $Q$), we recover the ordinary loss:
\begin{equation}
\label{eqn:unbias}
    \Ebb_{Q}\Big[\frac{1}{|\Dcal|} \sum_{(u,i)\in\Dcal} \frac{\phi\big(y_{u,i}\cdot f_{\theta}(\xbf_u, \zbf_i) \big)}{p(O_{u,i}=1 | \xbf_u, \zbf_i)} \Big] = \Ebb_{P_n} \Big[\frac{\phi\big(Y\cdot f_{\theta}(\Xbf, \Zbf) \big)}{p(O=1 |\Xbf, \Zbf)}p(O=1 |\Xbf, \Zbf) \Big] = \ell_{\phi}(f_{\theta}),
\end{equation}
where the second expectation is taken with respect to the empirical distribution $P_n$. Let $Q_0$ be the distribution for the underlying exposure mechanism. The propensity-weighted empirical distribution is then given by $P_n/Q_0$ (after scaling), which can be think of as the synthetic sample distribution after eliminating the influence from the underlying exposure mechanism. It is straightforward to verify that after scaling, the expected propensity-weighted loss is exactly given by: $\Ebb_{P_n/Q_0}\big[\phi(Y\cdot f_{\theta}(\Xbf, \Zbf)) \big]$.

\textbf{The hidden assumption of the missing-data (click) model}

A number of prior work deals with the unidentifiable exposure mechanism by assuming a missing-data model \cite{saito2020unbiased,ai2018unbiased,liang2016modeling,wang2018modeling}, which is also referred to as the \emph{click model}:
\begin{equation}
\label{eqn:click}
p(\text{click}=1|x)=p(\text{expose}=1|x)\cdot p(\text{relevance}=1|x).
\end{equation}
While the \emph{click model} greatly simplifies the problem since the exposure mechanism can now be characterized explicitly, it relies on a hidden assumption that is rarely satisfied in practice. We use $R$ to denote the relevance and $Y$ to denote the click. The fact that $Y=1 \Leftrightarrow O=1$ and $R=1$ implies:
\begin{equation*}
\begin{split}
    & \quad p(Y=1 | x) = p(O=1, R=1 | x) = p(O=1 | x) \cdot p(R=1 | O=1, x) \\
    & \overset{(\ref{eqn:click})}{\Longrightarrow} p(R=1 | O=1, x) = p(R=1 | x),
\end{split}
\end{equation*}
which suggests that being relevant is independent of getting exposed given the features. This is rarely true (or at least cannot be examined) in many real-world problems, unless $x$ contains every single factor that may affect the exposure and user preference.
We aim at providing a robust solution whenever the hidden assumption of the missing-data (click) model is dubious or violated. 

\section{Method}
\label{sec:method}
Let $P^*$ be the ideal exposure-eliminated sample distribution corresponding to $P/Q_0$, according to the underlying exposure mechanism $Q_0$ and data distribution $P$. For notation simplicity, without overloading the original meaning by too much, from this point we treat $P$, $P_n$, $Q_0$ and $P^*$ as distributions on the sample space $\Xcal$ which consists of all the observed data $(\xbf_u,\zbf_i, y_{u,i})$ with $(u,i)\in\Dcal$. Since we make no data or model assumptions that may allow us to accurately recover $P^*$, we introduce a minimax formulation to characterize the uncertainty. We optimize $f_{\theta}$ against the worst possible choice of (a hypothetical) $\hat{P}$, whose discrepancy with the ideal $P^*$ can only be determined by the data to a neighborhood: $\text{Dist}(P^*,\hat{P}) < \rho$. Among the divergence and distribution distance measures, we choose the Wasserstein distance for our problem, which is defined as:
\begin{equation}
\label{eqn:wasserstein}
W_c(\hat{P}, P^*) = \inf_{\gamma \in \Pi(\hat{P}, P^*)} \Ebb_{((\xbf, \zbf, y),(\xbf', \zbf', y'))\sim \gamma} \big[ c\big( (\xbf, \zbf, y), (\xbf', \zbf', y')  \big) \big],
\end{equation}
where $c: \Xcal \times \Xcal \to [0,+\infty)$ is the convex, lower semicontinuous transportation cost function with $c(\tbf, \tbf)=0$, and $\Pi(\hat{P}, P^*)$ is the set of all distributions whose marginals are given by $\hat{P}$ and $P^*$. Intuitively, the Wasserstein distance can be interpreted as the minimum cost associated with transporting mass between probability measures. We choose the Wasserstein distance instead of others exactly because we wish to understand how to transport from the empirical data distribution to an ideal synthetic data distribution where the observations were independent of the exposure mechanism. Hence, we consider the local minimax \emph{empirical risk minimization (ERM)} problem:
\begin{equation}
\label{eqn:minimax-1}
    \underset{f_{\theta} \in \mathcal{F}}{\text{minimize}} \sup_{W_c(P^*,\hat{P}) < \rho} \Ebb_{\hat{P}}\big[ \delta(Y, f_{\theta}(\Xbf, \Zbf)) \big],
\end{equation}
where we directly account for the uncertainty induced by the lack of identifiability in the exposure mechanism, and optimize $f_{\theta}$ under the worst possible setting. However, the formulation in (\ref{eqn:minimax-1}) is first of all a constraint optimization problem. Secondly, the constraint is expressed in terms of the hypothetical $P^*$. After applying a duality argument, we express the dual problem via the exposure mechanism in the following Claim \ref{claim:dual}. We use $\hat{Q}$ to denote some estimation of $Q_0$.

\begin{claim}
\label{claim:dual}
Suppose that  the transportation cost $c$ is continuous and the propensity score are all bounded away from zero, i.e. $p(O_{i,u}=1 | \xbf_u, \zbf_i) \geq \mu$. Let $\mathcal{P} = \{P: W_c(P^*,P) < \rho\}$, then
\begin{equation*}
\label{eqn:dual-form}
\begin{split}
    \sup_{\hat{P}\in \mathcal{P}} \Ebb_{\hat{P}}\big[ \delta(Y,  f_{\theta}(\Xbf, \Zbf)) \big] = \inf_{\alpha \geq 0} \Big\{ \alpha\rho + \sup_{\hat{Q}} \big\{ \Ebb_{P}\Big[ \frac{\delta(Y, f_{\theta}(\Xbf, \Zbf))}{\hat{q}(O=1|\Xbf, \Zbf)} \Big] - c_0 \alpha W_c(\hat{Q}\inv, Q_0\inv) \big\}  \Big\},
\end{split}
\end{equation*}
where $c_0$ is a positive constant and $\hat{q}$ is the density function associated with $\hat{Q}$.
\end{claim}
We defer the proof to Appendix \ref{append:claim2}. If we consider the relaxation for each fixed $\alpha$ (see the appendix), the minimax objective has a desirable formulation where $\alpha$ becomes a tuning parameter:
\begin{equation}
\label{eqn:minimax-2}
\underset{f_{\theta} \in \mathcal{F}}{\text{minimize}} \sup_{\hat{Q}} \Ebb_{P}\Big[ \frac{\delta(Y, f_{\theta}(\Xbf, \Zbf))}{\hat{q}(O=1|\Xbf, \Zbf)} \Big] - \alpha W_c(\hat{Q}, Q_0), \quad \alpha \geq 0.
\end{equation}
To make sense of (\ref{eqn:minimax-2}), we see that while $\hat{Q}$ is acting adversarially against $f_{\theta}$ as the inverse weights in the first term, it cannot arbitrarily increase the objective function, since the second terms acts as a regularizer that keeps $\hat{Q}$ close to the true exposure mechanism $Q_0$. Compared with the primal problem in (\ref{eqn:minimax-1}), the relaxed dual formulation in (\ref{eqn:minimax-2}) gives the desired unconstrained optimization problem. Also, we point out that the exposure mechanism is often given by the recommender system that was operating during the data collection, which we shall leverage as a domain knowledge to further convert (\ref{eqn:minimax-2}) to a more tractable formulation. Let $g^*$ be the recommendation model that underlies $Q_0$. Assume for now that $p_{g}(O=1|\Xbf, \Zbf)$ is given by $G\big(g(\Xbf, \Zbf)\big) \in (\mu,1), \mu>0$ for some transformation function $G$. We leave the inclusion and manipulation of the unobserved factors to Section \ref{sec:practical}. The objective in (\ref{eqn:minimax-2}) can then be converted to a two-model adversarial game:
\begin{equation}
\label{eqn:minimax-3}
\underset{f_{\theta} \in \mathcal{F}}{\text{minimize}} \sup_{g_{\psi} \in \Gcal} \Ebb_{P}\Big[ \frac{\delta(Y, f_{\theta}(\Xbf, \Zbf))}{G\big(g_{\psi}(\Xbf, \Zbf) \big)} \Big] - \alpha W_c(G(g_{\psi}), G(g^*)), \quad \alpha \geq 0.
\end{equation}

Before we move on to discuss the implications of (\ref{eqn:minimax-3}), its practical implementations and the minimax optimization, we first show and discuss the theoretical guarantees for the generalization error, in comparison to the standard ERM setting, after introducing the adversarial component.

\subsection{Theoretical property}
\label{sec:theory}

Before we state the main results, we need to characterize the loss function corresponding to the adversarial objective as well as the complexity of our hypothesis space. For the first purpose, we introduce the cost-regulated loss which is defined as:
$
\Delta_{\gamma}\big(f_{\theta}; (\xbf, \zbf, y)\big) = \sup_{(\xbf', \zbf', y')\in \Xcal} \Big\{\frac{\delta\big (y', f_{\theta}(\xbf', \zbf')\big)}{q(o=1|\xbf', \zbf')} - \gamma c\big( (\xbf, \zbf, y), (\xbf', \zbf', y')\big)  \Big\},
$
For the second purpose, we consider the \emph{entropy integral}
$
\Jcal(\tilde{\Fcal}) = \int_{0}^{\infty} \sqrt{\log \Ncal(\epsilon; \tilde{\Fcal},\|.\|_{\infty})} d \epsilon,
$
where $\tilde{\Fcal} = \{\delta(f_{\theta,\cdot}) | f_{\theta} \in \Fcal\}$ is the hypothesis class and $\Ncal(\epsilon; \tilde{\Fcal},\|\cdot \|_{\infty})$ gives the \emph{covering number} for the $\epsilon-$cover of $\tilde{\Fcal}$ in terms of the $\|\cdot \|_{\infty}$ norm. Suppose that $|\delta(y, f_{\theta}(\xbf, \zbf))| \leq M$ holds uniformly. Now we state our main theoretical result on the worst-case generalization bound under the minimax setting, and the proof is delegated to Appendix \ref{append:thm1}.

\begin{theorem}
\label{thm:generalization}
Suppose the mapping $G$ from $g_{\psi}$ to $q(o=1|\xbf, \zbf)$ is one-to-one and surjective with $g_{\psi} \in \Gcal$. Let $\tilde{\Gcal}(\rho) = \big\{g_{\psi} \in \Gcal \, | \, W_c\big(G(g_{\psi}), G(g^*)\big) \leq \rho \big\}$. Then under the conditions specified in Claim \ref{claim:dual}, for all $\gamma \geq 0$ and $\rho > 0$, the following inequality holds with probability at least $1-\epsilon$:
\begin{equation*}
 \sup_{g_{\psi} \in \tilde{\Gcal}(\rho)}  \Ebb_{P} \Big[ \frac{\delta(Y, f_{\theta}(\Xbf, \Zbf))}{G\big(g_{\psi}(\Xbf, \Zbf)\big)} \Big] \leq  c_1\gamma \rho + \Ebb_{P_n}[\Delta_{\gamma}\big(f_{\theta}; (\Xbf, \Zbf, Y)\big)] + \frac{24\Jcal(\tilde{\Fcal}) + c_2(M, \sqrt{\log \frac{2}{\epsilon}}, \gamma)}{\sqrt{n}},
\end{equation*}
where $c_1$ is a positive constants and $c_2$ is a simple linear function with positive weights.
\end{theorem}
The above generalization bound holds for all $\rho$ and $\delta$, and we show that when they are decided by some data-dependent quantities, the result can be converted to some simplified forms that reveal the more direct connections with the propensity-weighted loss and standard ERM results (with the proof provided in Appendix \ref{append:corollary1}).
\begin{corollary}
\label{corollary:one}
Following the statements in Theorem \ref{thm:generalization}, there exists some data-dependent $\gamma_n$ and $\rho_n(f_{\theta})$, such that
when $\gamma \geq \gamma_n$, for all $\rho>0$:
\begin{equation*}
    Pr\Big(\sup_{g_{\psi} \in \tilde{\Gcal}(\rho)}  \Ebb_{P} \Big[ \frac{\delta(Y, f_{\theta}(\Xbf, \Zbf))}{G\big(g_{\psi}(\Xbf, \Zbf)\big)} \Big] \leq  c_1\gamma \rho + \Ebb_{P_n}\Big[\frac{\delta\big(f_{\theta}; (\Xbf, \Zbf, Y)\big)}{q(O=1 | \Xbf, \Zbf)}\Big] + \varepsilon_n(\epsilon)  \Big) > 1-\epsilon;
\end{equation*}
and when $\rho = \rho_n(f_{\theta})$, for all $\gamma \geq 0$:
\begin{equation*}
    Pr\Big(\sup_{g_{\psi} \in \tilde{\Gcal}(\rho)}  \Ebb_{P} \Big[ \frac{\delta(Y, f_{\theta}(\Xbf, \Zbf))}{G\big(g_{\psi}(\Xbf, \Zbf)\big)} \Big] \leq  \sup_{P:W_c(P,P_n)\leq \tilde{\rho}}\Ebb_{P}\Big[\frac{\delta\big(f_{\theta}; (\Xbf, \Zbf, Y)\big)}{q(O=1 | \Xbf, \Zbf)}\Big] + \varepsilon_n(\epsilon)  \Big) > 1-\epsilon,
\end{equation*}
where $\varepsilon_n(\epsilon) = \big(24\Jcal(\tilde{\Fcal}) + c_2(M, \sqrt{\log \frac{2}{\epsilon}}, \gamma)\big)/\sqrt{n}$ as suggested by Theorem \ref{thm:generalization}.
\end{corollary}
Corollary \ref{corollary:one} shows that the proposed approach has the same $1/\sqrt{n}$ rate as the standard ERM. Also, the first result reveals an extra $\delta\rho$ bias term induced by the adversarial setting, the second result characterizes how the additional uncertainty is reflected on the propensity-weighted empirical loss.


    

\subsection{Practical implementations}
\label{sec:practical}

Directly optimizing the minimax objective in (\ref{eqn:minimax-3}) is infeasible since $g^*$ is unknown and the Wasserstein distance is hard to compute when $\Gcal$ is a complicated model such as neural network \cite{panaretos2019statistical}. Nevertheless, understanding the comparative roles of $f_{\theta}$ and $g_{\psi}$ can help us construct practical solutions. 

Recall that our goal is to optimize $f_{\theta}$. The auxiliary $g_{\psi}$ is introduced to characterize the adversarial exposure mechanism, so we are less interested in recovering the true $g^*$. With that being said, the term $W_c(G(g_{\psi}), G(g^*))$ only serves to establish certain regularizations on $g_{\psi}$ such that it is constrained by the underlying exposure mechanism. Relaxing or tightening the regularization term should not significantly impact the solution since we can always adjust the regularization parameter $\alpha$.
Hence, we are motivated to design tractable regularizers to approximate or even replace $W_c(G(g_{\psi}), G(g^*))$, as long as the constraint on $g_{\psi}$ is established under the same principle. Similar ideas have also been applied to train the \emph{generative adversarial network (GAN)}: the optimal classifier depends on the unknown data distribution, so in practice, people use alternative tractable classifiers that fit into the problem \cite{goodfellow2016nips}.
We list several alternative regularizers for $g_{\psi}$ as below.
\begin{itemize}
    \item In the explicit feedback data setting, the exposure status is partially observed, so the loss of $G(g_{\psi})$ on the partially-observed exposure data can be used as the regularizer, i.e. $\frac{1}{|\Dcal_{\text{exp}}|}\sum_{(u,i)\in \Dcal_{\text{exp}}}\phi\big(g_{\psi}(\xbf_u, \zbf_i)\big)$, where $\Dcal_{\text{exp}} = \{(u,i)\in \Dcal | o_{u,i}=1\}$.
    \item For the content-based recommendations, the exposure often have high correlation with popularity where the popular items are more likely to be recommended. So the regularizer may leverage the empirical popularity via: $\text{corr} \big(\frac{1}{m}\sum_{u}G(g_{\psi}(\Xbf_u, \Zbf_i)),  \frac{1}{m}\sum_{u}Y_{u,i} \big)$.
    \item In the implicit feedback setting, if all the other choices are impractical, we may simply use the loss on the feedback data as a regularizer: $\Ebb_{P_n}\big[ \phi(Y\cdot g_{\psi}(\Xbf, \Zbf)) \big]$. The loss-based regularizer is  meaningful because $g^*$ is often determined by some other recommendation models. If it happens that $g^* \in \Gcal$, we can expect similar performances from $g_{\psi}$ and $g^*$ on the same feedback data since the exposure mechanism is determined by $g^*$ itself. 
\end{itemize}
We focus on the third example because it applies to almost all cases without requiring excessive assumptions. Therefore, the practical adversarial objective is now given by:
\begin{equation}
\label{eqn:minimax-4}
\underset{f_{\theta} \in \mathcal{F}}{\text{minimize}} \sup_{g_{\psi} \in \Gcal} \Ebb_{P_n}\Big[ \frac{\delta(Y, f_{\theta}(\Xbf, \Zbf))}{G\big(g_{\psi}(\Xbf, \Zbf) \big)} \Big] - \alpha \Ebb_{P_n}\big[ \delta(Y, g_{\psi}(\Xbf, \Zbf)) \big]  , \quad \alpha \geq 0.
\end{equation}

In the next step, we study how to handle the unobserved factors who also play a part in the exposure mechanism. As we mentioned in Section \ref{sec:intro}, having unobserved factors is inevitable practically. In particular,
we leverage the \emph{Tukey's factorization} proposed in the missing data literature \cite{franks2016non}. In the presence of unobserved factors, Tukey's factorization suggests that we additionally characterize the relationship between exposure mechanism and outcome \cite{franks2019flexible} (see the appendix for detailed discussions). Relating the outcome to exposure mechanism has also been found in the recommendation literature \cite{schnabel2016recommendations}.
For clarity, we employ a simple logistic-regression to model $G$ as: 
\[
G_{\beta}\big(g_{\psi}(\xbf, \zbf), y\big) = \sigma \big(\beta_0 + \beta_1 g_{\psi}(\xbf, \zbf) + \beta_2 y  \big),
\]
where $\sigma(\cdot)$ is the sigmoid function. We now reach the final form of the adversarial game:
\begin{equation}
\label{eqn:minimax-5}
\underset{f_{\theta} \in \mathcal{F}, \beta}{\text{minimize}} \sup_{g_{\psi} \in \Gcal} \Ebb_{P_n}\Big[ \frac{\delta(Y, f_{\theta}(\Xbf, \Zbf))}{G_{\beta}\big(g_{\psi}(\Xbf, \Zbf), Y \big)} \Big] - \alpha \Ebb_{P_n}\big[ \delta(Y, g_{\psi}(\Xbf, \Zbf)) \big]  , \quad \alpha \geq 0.
\end{equation}
We place $\beta$ to the minimization problem for the following reason. By our design, $G_{\beta}$ merely characterizes the potential impact of unobserved factors which we do not consider to act adversarially. Otherwise, the adversarial model can be too strong for $f_{\theta}$ to learn anything useful.

\subsection{Minimax optimization and robust evaluation}
\label{sec:optimization}

\begin{wrapfigure}[11]{L}{0.4\textwidth} 
      \begin{algorithm}[H]   
      \label{algo:one}
        \SetCustomAlgoRuledWidth{0.38\textwidth}  
        \caption{Minimax optimization}
        \KwIn{Learning rates $r_{\theta}$, $r_{\psi}$, discounts $d_{\theta}, d_{\psi} > 1$;}
        \While{loss not stabilized}{
        $\theta = \theta - r_{\theta}\Ebb_{\text{batch}} \nabla_{\theta}\ell \big(f_{\theta}, g_{\psi}\big)$\;
        $\psi = \psi + r_{\psi}\Ebb_{\text{batch}} \nabla_{\psi}\ell \big(f_{\theta}, g_{\psi}\big)$\;
        $r_{\theta} = r_{\theta}/d_{\theta}$, $r_{\psi} = r_{\psi}/d_{\psi}$\;
        }
      \end{algorithm}
\end{wrapfigure}

To handle the adversarial training, we adopt the sequential optimization setup where the players take turn to update their model.
Without loss of generality, we treat the objective in (\ref{eqn:minimax-4}) as a function of of the two models: $\min_{f_{\theta}} \max_{g_{\psi}}\ell \big(f_{\theta}, g_{\psi}\big)$. When $\ell$ is nonconvex-nonconcave, the classical Minimax Theorem no longer hold and $\min_{f_{\theta}} \max_{g_{\psi}}\ell \big(f_{\theta}, g_{\psi}\big) \neq \max_{g_{\psi}} \min_{f_{\theta}} \ell \big(f_{\theta}, g_{\psi}\big)$  \cite{terkelsen1973some}. Consequently, which player goes first has important implications. 
Here, we choose to train $f_{\theta}$ first because $g_{\psi}$ can then choose the worst candidate from the uncertainty set in order to undermines $f_{\theta}$.
We adopt the two-timescale gradient descent ascent (GDA) \cite{heusel2017gans} schema that is widely applied to train adversarial objectives (Algorithm \ref{algo:one}). However, the existing analysis on GDA's converging to local Nash equilibrium assumes simultaneous training \cite{heusel2017gans,ratliff2013characterization,prasad2015two}, so their guarantees do not apply here. Instead, we keep training until the objective stops changing by updating either $f_{\theta}$ or $g_{\psi}$. 

Consequently, the stationary points in Algorithm \ref{algo:one} may not attain local Nash equilibrium. Nevertheless, when the timescale of the two models differ significantly (by adjusting the initial learning rates and discounts), it has been shown that the stationary points belong to the \emph{local minimax solution} up to some degenerate cases \cite{jin2019local}. The local minimaxity captures the \emph{optimal strategies} in the sequential game if both
models are only allowed to change their strategies locally. Hence, Algorithm \ref{algo:one} leads to solutions that are locally optimal. Finally, the role of $G_{\beta}$ is less important in the sequential game, and we do not observe significant differences from updating it before or after $f_{\theta}$ and $g_{\psi}$.


Recommenders are often evaluated by the \emph{mean square error (MSE)} on explicit feedback, and by the information retrieval metric such as DCG and NDCG on implicit feedback. 
After the training, we obtain the candidate model $f_{\theta}$ as well as the $G_{\beta}(g_{\psi})$ who gives the worst-case propensity score function specialized for $f_{\theta}$. Therefore, instead of pursuing unbiased evaluation, we instead consider the \emph{robust evaluation} by using $G_{\beta}(g_{\psi})$. It frees the offline evaluation from the potential impact of exposure mechanism, and thus provide a robust view on the true performance. For instance, the robust NDCG can be computed via: $\frac{1}{|\Dcal_{\text{test}}|} \sum_{(u,i)\in \Dcal_{\text{test}}} \text{NDCG}\big(y_{u,i}, f_{\theta}(\xbf_u, \zbf_i)\big) \big/ G_{\beta}\big(g_{\psi}(\xbf_u, \zbf_i)\big)$.

\section{Relation to other work}
\label{sec:related}
The propensity-weighting method is proposed and intensively studied in the observation studies and causal inference literature \cite{austin2011introduction,austin2015moving}. A recent work that introduces adversarial training to solve the identifiability issue studies on the covariate-balancing methods \cite{kallus2018deepmatch,yoon2018ganite}. Adversarial training is widely applied by such as generative models \cite{goodfellow2016nips}, model defense \cite{tramer2017ensemble}, adversarial robustness \cite{xie2019feature} and distributional robust optimization (DRO) \cite{rahimian2019distributionally}. Compare with GAN, we study the sampling distribution instead of the generating distribution, and GAN does not involve counterfactual modelling. DRO often focus on the feature distribution while we study the propensity score distribution. Using the Wasserstein distance as regularization is also common in the literature \cite{shafieezadeh2019regularization,gao2017wasserstein}. Here, we introduce the adversarial setting for the identifiability issue, whereas the model defense and adversarial robustness study the training and modelling properties under deliberate adversarial behaviors. 

Counterfactual modelling for recommenders often relies on certain data or model assumptions (such as the click model assumption) to make up for the identifiability issue, and is thus venerable when the assumptions are violated in practice \cite{schnabel2016recommendations,liang2016causal,joachims2017unbiased,agarwal2018counterfactual,liang2016modeling,yang2018unbiased,hernandez2014probabilistic,saito2020unbiased,ai2018unbiased,wang2018modeling}. Adversarial training for recommenders often borrows the GAN setting by assuming a generative distribution for certain components \cite{wang2017irgan,he2018adversarial}. Here, we do not assume the generative nature of recommender systems.

\section{Experiment and Result}
\label{sec:experiment}
We conduct simulation study, real-data analysis, as well as online experiments to demonstrate the various benefits of the proposed adversarial counterfactual learning and evaluation approach.  
\begin{itemize}
    \item In the simulation study, we generate the synthetic data using real-world explicit feedback dataset so that we have access to the oracle exposure mechanism. We then show that models trained by our approach achieve superior unbiased offline evaluation performances.
    \item In the real-world data analysis, we demonstrate that the models trained by our approach also achieve more improvements even using the standard offline evaluation.
    \item By conducting online experiments, we verify that our robust evaluation is more accurate than the standard offline evaluation when compared with the actual online evaluations.
\end{itemize}

As for the baseline models, since we are proposing a high-level learning and evaluation approach that are compatible with almost all the existing recommendation models, we consider the well-known baseline models to demonstrate the effectiveness of our approach. Specifically, we employ the popularity-based recommendation \textbf{(Pop)}, matrix factorization collaborative filtering \textbf{(CF)}, multi-layer perceptron-based CF model \textbf{(MLP)}, neural CF \textbf{(NCF)} and the generalized matrix factorization \textbf{(GMF)}, as the representatives for the content-based recommendation. We also consider the prevailing attention-based model \textbf{(Attn)} as a representative for the sequential recommendation. We also choose $f_{\theta}$ and $g_{\psi}$ among the above baselines models for our adversarial counterfactual learning. To fully demonstrate the effectiveness of the proposed adversarial training, we also experiment with the non-adversarially trained propensity-score method \textbf{PS}, where we first optimize $g_{\psi}$ only on the regularization term until convergence, keep it fixed, and then train $f_{\theta}$ in the regular propensity-weighted ERM setting. For the sake of notation, we refer to our learning approach as the \textbf{ACL-}.

We choose to examine the various methods with the widely-adopted next-item recommendation task. In particular, all but the last two user-item interactions are used for training, the second-to-last interaction is used for validation, and the last interaction is used for testing. All the detailed data processing, experiment setup, model configuration, parameter tuning, training procedure, validation, testing and sensitivity analysis are provided in Appendix \ref{append:experiment}.

\begin{table}[!]
    \centering
    \resizebox{\columnwidth}{!}{%
    \begin{tabular}{c c c c c c c c c c c}
        \hline \hline
       & MLP & MLP & MLP & GMF & GMF & GMF & \textbf{ACL-MLP} & \textbf{ACL-GMF} \\
      \hdashline
       \textbf{config} & Pop & MLP & Oracle & Pop & GMF & Oracle & MLP & GMF \\
      \hline
        \multicolumn{9}{c}{\textbf{\textit{MovieLens-1M}}} \\
        Hit@10 &  39.60 (.12 & 39.24 (.3) & \underline{39.68} (.3) & 39.00 (.2) & 39.47 (.1) & 39.10 (.2) & \textbf{40.32} (.1) & 39.08 (.2) \\
        NDCG@10 & 20.26 (.1) & 20.10 (.2) & \underline{20.33} (.2) & 19.33 (.3) & 19.58 (.2) & 19.30 (.1) & \textbf{20.81} (.2) & 19.61 (.1) &\\
        \hline
        \multicolumn{9}{c}{\textbf{\textit{Goodreads}}} \\
        Hit@10 & 31.90 (.3) & 30.61 (.2) & \textbf{33.82} (.1) & 30.01 (.3) & 31.36 (.2) & 33.50 (.1) & \underline{33.45} (.2) & 32.51 (.2)\\
        NDCG@10 & 16.65 (.2) & 15.72 (.2) & \textbf{17.81} (.1) & 15.24 (.2) & 16.40 (.2) & 17.28 (.2) & \underline{17.50} (.1) & 16.85 (.1)\\
        \hline \hline
        \multicolumn{9}{c}{\vspace{0.5cm}} \\
        \hline \hline
         Data& \multicolumn{4}{c}{\textbf{\textit{MovieLens-1M}}} & \multicolumn{4}{c}{\textbf{\textit{Goodreads}}} \\
         \hline
        Model & Pop & CF & MLP & GMF & Pop & CF & MLP & GMF \\ \hline 
        Hit@10 & 33.76 (.1) & 38.27 (.2) & 39.43 (.2) & 39.00 (.2) & 26.62 (.3) & 30.90 (.2) & 31.78 (.2) & 29.59 (.3)\\
        NDCG@10 & 17.75 (.1) & 18.59 (.2) & 20.09 (.2) & 19.28 (.3) & 14.29 (.2) &  16.43 (.1) & 16.58 (.2) & 14.94 (.2) \\ \hline \hline \\
    \end{tabular}
    }
    \caption{\small Unbiased evaluations (using the true exposure) for the baselines and the proposed approach on the semi-synthetic data. \textbf{Upper panel:} we provide in the \textbf{config} rows the $g_{\psi}$ model (such as using the baseline models and the oracle model) when trained with the propensity-score (PS) approach or the proposed approach (marked by the \textbf{ACL-}). \textbf{Lower panel:} the original baseline models without using propensity-score approach or ACL. We use bold-font and underscore to mark the best and second-best outcomes. The mean and standard deviation are computed over ten repetitions, and the complete numerical results are deferred to Appendix \ref{append:experiment}.}
    \label{tab:synthetic}
\end{table}

\begin{table}[!]
    \centering
    \resizebox{\columnwidth}{!}{%
    \begin{tabular}{c c c c c c c c c c}
    \hline \hline
         & Pop & CF & MLP & NCF & GMF & Attn & PS & \textbf{ACL} & \textbf{ACL} \\
        \hline
        \multicolumn{10}{c}{\textbf{\textit{MovieLens-1M}}} \\
        \hline
        \textbf{config} & & & & & & & Attn / Pop & GMF / GMF & Attn / Attn \\
        \hdashline
        Hit@10 & 42.18 (.2) & 60.97 (.1) & 61.01 (.2) & 63.37 (.3) & 63.97(.1) & 82.66 (.2) & 81.97 (.1) & 64.32 (.2) & \textbf{83.64} (.1) \\
        NDCG@10 & 21.99 (.1) & 32.59 (.1) & 32.09 (.3) & 33.49 (.1) & 33.82(.2) & 55.27 (.1) & 54.51 (.1) & 33.70 (.1) & \textbf{55.71} (.2) \\
        \hline
        \multicolumn{10}{c}{\textbf{\textit{LastFM}}} \\
        \hline
        \textbf{config} & & & & & & & GMF / Pop & GMF / GMF & Attn /Attn \\
        \hdashline
        Hit@10 &  25.26 (.2) & 52.97 (.3) & 81.86 (.3) & 81.87 (.3) & 83.12 (.3) & 71.89 (.3) & 82.64 (.2) & \textbf{83.64} (.2) & 72.02 (.2) \\
        NDCG@10 & 15.35 (.1) & 31.54 (.2) & 58.38 (.2) & 57.33 (.4) & 58.96 (.2) & 59.75 (.2) & 58.84 (.2) & 59.11 (.1) & \textbf{59.45} (.1)\\
        \hline
        \multicolumn{10}{c}{\textbf{\textit{Goodreads}}} \\
        \hline
        \textbf{config} & & & & & & & Attn / Pop & GMF / GMF & Attn / Attn \\
        \hdashline
        Hit@10 & 43.36 (.1) & 60.32 (.2) & 62.17 (.2) & 63.11 (.3) & 63.78 (.1) & 72.63 (.2) & 73.39 (.1) & 64.17 (.2) & \textbf{73.82} (.3)\\
        NDCG@10 & 22.73 (.2) & 37.73 (.1) & 37.65 (.1) & 38.78 (.3) & 38.69 (.1) & 48.98 (.1 ) & 49.92 (.3) & 39.53 (.1) & \textbf{49.99} (.1) \\
        \hline \hline \\
    \end{tabular}
    }
    \caption{\small Standard evaluations (without accounting for exposure) for the baselines and proposed approach on the benchmark data. Similarly, we provide in the \textbf{config} rows the $f_{\theta}$ and $g_{\psi}$ model choice when trained with the PS and our ACL approach. We present here the best $f_{\theta}$ and $g_{\psi}$ combination for the PS method, and the full results for our approach and the baselines are deferred to Appendix \ref{append:experiment}.}
    \label{tab:real}
\end{table}

\textbf{Synthetic data analysis}. 
We use the explicit feedback data from \emph{MovieLens-1M}\footnote{All the data sources, processing steps and other detailed descriptions are provided in the appendix.} and \emph{Goodreads} datasets. We train a baseline CF model and use the optimized hidden factors to generate a synthetic exposure mechanism (with the details presented in Appendix \ref{append:simulation}), and treat it as the \emph{oracle} exposure. The implicit feedback data are then generated according to the oracle exposure as well as the optimized hidden factors. Unbiased offline evaluation 
is now possible because we have access to the exposure mechanism. Also, to set a reasonable benchmark under our simulation setting, we provide the additional experiments where $g_{\psi}$ is given by the oracle exposure model. The results are provided in Table \ref{tab:synthetic}. We see that when trained with the proposed approach, the baselines models yield their best performances (other than the oracle-enhanced counterparts) under the unbiased offline evaluation, and outperforms the rest of the baselines, which reveals the first appeal of our approach.

\textbf{Real data analysis.} 
Other than using the \emph{MovieLens-1M} and \emph{Goodreads} data in the implicit feedback setting, we further include the \emph{LastFM} music recommendation (implicit feedback) dataset. From the results in Table \ref{tab:real}, we observe that the models trained by our approach achieve the best outcome, even using the standard evaluation where the exposure mechanism is not considered. The better performance in standard evaluation suggests the second appeal of the adversarial counterfactual learning, that even though it optimizes towards the minimax setting, the robustness is not at the cost of the performance under the standard evaluation.

\begin{table}[hbt]
\centering
    \resizebox{0.8\columnwidth}{!}{
      \begin{tabular}{c| c c c c c} \hline \hline
    MSE on metric & Standard & Popularity debiased & Propensity model debiased & Robust \\ \hline 
    Hit@5 & .18(.10) & .14(.08) &  .14(.06) & \textbf{.12}(.04) \\ \hline
    NDCG@5 & .10(.06) & .09(.05) & .08(.05) & \textbf{.07}(.03) \\ \hline \\
    \end{tabular}
    }
    \caption{\small The mean-squared error (MSE) to online evaluation results from eight online experiments.}
    \label{tab:online_exp}
  \end{table}

\textbf{Online experiment analysis.} 
To examine the practical benefits of the proposed robust learning and evaluation approach in real-world experiments, we carry out several online A/B testings on the \textit{Walmart.com}, a major e-commerce platform in the U.S., in a content-based item recommendation setting. We are provided with the actual online testing and evaluation results. All the candidate models were trained offline using the proposed approach. We compare the standard offline evaluation, popularity-debiased offline evaluation (where the item popularity is used as the propensity score), the propensity-score model approach and our robust evaluation, with respect to the actual online evaluations. In Table \ref{tab:online_exp}, we see that our proposed evaluation approach is indeed a more robust approximation to the online evaluation. It reveals the third appeal of the proposed approaches that they are capable of narrowing the gap between online and offline evaluations.

        


\section{Conclusion}
\label{sec:conclusion}
We thoroughly analyze the drawback of supervised learning for recommender systems and propose the theoretically-grounded adversarial counterfactual learning and evaluation framework. We provide elaborated theoretical and empirical results to illustrate the benefits of the proposed approach. \\
\textbf{Scope and limitation}. The improvement brought by our approach ultimately depends on the properties of the feedback data, e.g. to what extent is the identifiability issue causing uncertainties in the data. Also, we observe empirically that the propensity model can experience undesired behaviors during the adversarial training as a consequence of using suboptimal tuning parameters. Therefore, it remains to be studied how the optimization dynamics can impact the two-model interactions for the proposed adversarial counterfactual learning.

\section*{Broader Impact}
\label{sec:impact}
To the best of our knowledge, the approaches discussed in this paper raise no major ethical concerns and societal consequences. Researchers and practitioners from the recommender system domain may benefit from our research since robust offline learning and evaluation has been a significant challenge in real-world applications. The worst possible outcome when the proposed approach fails is that it reduces to the standard offline learning as the propensity model stops making the desired impact. Finally, the proposed approach aims at solving the identifiability issues of the data, the extent of which depends on the properties of the data.

\begin{ack}
The work is supported by the Walmart U.S. eCommerce. The authors declare that there is no conflict of interest.
\end{ack}

\bibliographystyle{abbrvnat}
\bibliography{references}

\clearpage
\appendix

\newcommand{\Hcal}{\mathcal{H}}
\newcommand{\hbf}{\mathbf{h}}
\newcommand{\Hbf}{\mathbf{H}}

\setcounter{equation}{0}
\renewcommand{\theequation}{A.\arabic{equation}}
\setcounter{figure}{0}
\renewcommand{\thefigure}{A.\arabic{figure}}
\setcounter{table}{0}
\renewcommand{\thetable}{A.\arabic{figure}}
\setcounter{section}{0}
\renewcommand{\thesection}{A.\arabic{section}}


%



{\LARGE{\textbf{Appendix}}}
\\

We provide in the Appendix proofs for the major theoretical results. We also discuss the relaxation for Claim \ref{claim:dual}, and the origins as well as the implications from the Tukey's factorization on unobserved factors, which leads to our final two-model adversarial objective in (\ref{eqn:minimax-5}). 

We also describe the experiment details and provide the complete numerical results, including demonstrations that reveal the adversarial training process of the two models.

\section{Proof for Claim 1}
\label{append:claim1}

\begin{proof}
When taking the exposure mechanism into account, minimizing $f_{\theta}$ over the loss is implicitly doing
$
 \inf_{f_{\theta}}L_{\phi}\big(f_{\theta}, \{p^{(1)}, p^{(-1)}\}\big)
$,
where 
\begin{equation*}
\begin{split}
     L_{\phi}\big(f_{\theta}, \{p^{(1)}, p^{(-1)}\}\big) &= \Ebb \Big[ \phi\big( Y\cdot f_{\theta}(\xbf, \zbf; O)  \big)\Big] \\
    & = \sum_{o\in\{0,1\}} \phi\big(  f_{\theta}(\xbf, \zbf; O=o)\big) p^{(1)}(o) + \phi\big( - f_{\theta}(\xbf, \zbf; O=o)\big) p^{(-1)}(o).
\end{split}
\end{equation*}
For any fixed exposure mechanism $p(O|\xbf, \zbf)$, we have
\begin{equation}
\label{eqn:claim1}
\begin{split}
\inf_{f_{\theta}}L_{\phi}\big(f_{\theta}, \{p^{(1)}, p^{(-1)}\}\big) &= \sum_{o\in\{0,1\}} \inf_{\alpha}\big\{ \phi(\alpha)p^{(1)}(o) + \phi(-\alpha)p^{(-1)}(o) \big\} \\
& = \sum_{o\in\{0,1\}} p^{(1)}(o)\inf_{\alpha} \Big\{\phi(\alpha) + \phi(-\alpha)\frac{p^{(-1)}(o)}{p^{(1)}(o)} \Big\}.
\end{split}
\end{equation}
For each $o\in\{0,1\}$, let $\mu(o) = p^{(-1)}(o) / p^{(1)}(o)$ and $\Delta(\mu) = -\inf_{\alpha}\big( \phi(\alpha) + \phi(-\alpha \mu)\big)$. 

Notice that $\Delta(\mu)$ is a convex function of $\mu$ since the supremum (negative of the infimum) over a set of affine functions is convex. Since $\Delta$ is convex and continuous, we get:
\[
\inf_{f_{\theta}}L_{\phi}\big(f_{\theta}, \{p^{(1)}, p^{(-1)}\}\big) = -\sum_{o\in\{0,1\}} p^{(1)}(o) \Delta\Big(\frac{p^{(-1)}(o)}{p^{(1)}(o)} \Big),
\] which is exactly the f-divergence $D_{\Delta}(P^{(1)} || P^{(-1)})$ induced by $\Delta$. 

Also, up on achieving the infimum in (\ref{eqn:claim1}), the optimal $f_{\theta}$ is given by solving $a^*_{\phi}(\mu) = \arg\min_{\alpha} \big(\phi(\alpha) + \phi(-\alpha)\mu \big)$.
\end{proof}

\section{Proof for Claim 2 and the relaxation}
\label{append:claim2}
We first proved the dual formulation for the minimax ERM stated in Claim 2, and then discuss the relaxation for the dual problem.

\begin{proof}
For the estimation $\hat{P} = P/\hat{Q}$ of the ideal exposure-eliminated sample, $W_c(\hat{P},P^*)\leq \rho$ is equivalent to $W_c\big(P/\hat{Q},P/Q_0\big) \leq \rho$. 

The key observation is that when $P$ is given by the empirical distribution that assigns uniform weights to all samples, the Wasserstein's distance $W_c\big(P/\hat{Q},P/Q_0\big)$ is convex in $\hat{Q}\inv$ (since $c$ is convex) and $\hat{Q}=Q_0$ gives $W_c\big(P/\hat{Q},P/Q_0\big)=0$. 

Since we assume that the propensity scores are all bounded away from zero, so $P/\hat{Q}$ and $P/Q_0$ exist and and have normal behavior. So we able to establish the duality results, since the Slater's condition holds. Let $\hbf = (\xbf, \zbf, y) \in \Xcal$ and $\Xcal'$ be a copy of $\Xcal$. We have:

\begin{equation}
\label{eqn:claim-2-0}
\begin{split}
    & \sup_{\hat{P}:W_c(\hat{P},P^*)\leq \rho} \int \delta\big(y,f_{\theta}(\xbf, \zbf)\big) d\hat{P}(\hbf) \\
    & = \sup_{\hat{Q}:W_c\big(P/\hat{Q},P/Q_0\big)\leq \rho} \int \frac{\delta\big(y,f_{\theta}(\xbf, \zbf)\big)}{\hat{q}(O=1 \,|\, \xbf, \zbf)} d\hat{Q}(\hbf) \\
    &  = \inf_{\alpha\geq 0} \sup_{\hat{Q}} \Big\{ \int \frac{\delta\big(y,f_{\theta}(\xbf, \zbf)\big)}{\hat{q}(O=1 \,|\, \xbf, \zbf)} d\hat{Q}(\hbf) - \alpha W_c\big(P/\hat{Q},P/Q_0\big) + \alpha \rho \Big\} \\
    & = \inf_{\alpha\geq 0} \sup_{\hat{Q}} \Big\{ \int \frac{\delta\big(y,f_{\theta}(\xbf, \zbf)\big)}{\hat{q}(O=1 \,|\, \xbf, \zbf)} d\hat{Q}(\hbf) - \alpha \inf_{\gamma \in \Pi\big(P/\hat{Q}, P/Q_0\big)}\int c(\hbf, \hbf')d\gamma(\hbf, \hbf') + \alpha \rho \Big\} \\
    & \overset{}{=} \inf_{\alpha\geq 0} \sup_{\hat{Q}} \sup_{\gamma \in \Pi\big(P/\hat{Q}, P/Q_0\big)} \Big\{ \int \big( \frac{\delta_{f_{\theta}}(\hbf)}{\hat{q}(\hbf)} - \alpha c(\hbf, \hbf')\big)d\gamma(\hbf, \hbf') + \alpha \rho   \Big\} ,
\end{split}
\end{equation}
where in the last line we use the shorthand notation $\delta_{f_{\theta}}(\hbf):= \delta\big(y,f_{\theta}(\xbf, \zbf)\big)$ and $\hat{q}(\hbf) := \hat{q}(O=1|\xbf,\zbf)$. Then notice that
\begin{equation}
\label{eqn:claim-2-1}
    \sup_{\hat{Q}} \sup_{\gamma \in \Pi\big(\frac{P}{\hat{Q}}, \frac{P}{Q_0}\big)}  \int \Big( \frac{\delta_{f_{\theta}}(\hbf)}{\hat{q}(\hbf)} - \alpha c(\hbf, \hbf')\Big)d\gamma(\hbf, \hbf') \leq \int \sup_{\hbf \in \Xcal}\Big(\frac{\delta_{f_{\theta}}(\hbf)}{\hat{q}(\hbf)} - \alpha c(\hbf, \hbf')\Big)dQ_0(\hbf'),
\end{equation}
and we then show that the opposite direction also holds so it is always equality.
Let $\mathcal{K}$ be the space of measurable conditional distributions (Markov kernels) from $\Xcal$ to $\Xcal'$, then 
\begin{equation}
\label{eqn:claim-2-2}
\begin{split}
    \sup_{\hat{Q}} \sup_{\gamma \in \Pi\big(\frac{P}{\hat{Q}}, \frac{P}{Q_0}\big)}  \int \Big( \frac{\delta_{f_{\theta}}(\hbf)}{\hat{q}(\hbf)} &- \alpha c(\hbf, \hbf')\Big)d\gamma(\hbf, \hbf') \\ 
   &  \geq \sup_{K \in \mathcal{K}}\int  \Big(\frac{\delta_{f_{\theta}}(\hbf)}{\hat{q}(\hbf)} - \alpha c(\hbf, \hbf')\Big) dK(\hbf \, | \, \hbf') dQ_0(\hbf').
\end{split}
\end{equation}
In the next step, we consider the space of all measurable mappings $\hbf' \mapsto \hbf(\hbf')$ from $\Xcal'$ to $\Xcal$, denoted by $\Hcal$. Since all the mappings are measurable, the underlying spaces are regular, and $\delta_{f_{\theta}}$ and $c$ are at least semi-continuous, using standard measure theory arguments for exchanging the integration and supremum, we get
\begin{equation}
\label{eqn:claim-2-3}
    \sup_{\hbf(\cdot) \in \Hcal} \int \Big( \frac{\delta_{f_{\theta}}\big(\hbf(\hbf')\big)}{\hat{q}\big(\hbf(\hbf')\big)} - \alpha c\big(\hbf(\hbf'), \hbf'\big) \Big) dQ_0(\hbf') = \int \sup_{\hbf \in \Xcal} \Big( \frac{\delta_{f_{\theta}}(\hbf)}{\hat{q}(\hbf)} - \alpha c\big(\hbf, \hbf'\big) \Big) dQ_0(\hbf'),
\end{equation}
where the $\hbf(\cdot)$ on the LHS represents the mapping, and the $\hbf$ on the RHS still denotes elements from the sample space $\Xcal$.
Now we let the support of the conditional distribution $K(\hbf \,|\, \hbf')$ given by $\hbf(\hbf')$. So according to (\ref{eqn:claim-2-3}), we have:
\begin{equation}
\label{eqn:claim-2-4}
\begin{split}
\sup_{K \in \mathcal{K}}\int  \Big(\frac{\delta_{f_{\theta}}(\hbf)}{\hat{q}(\hbf)} &- \alpha c(\hbf, \hbf')\Big) dK(\hbf \, | \, \hbf') dQ_0(\hbf') \\ 
& = \sup_{\hbf(\cdot) \in \Hcal} \int \Big( \frac{\delta_{f_{\theta}}\big(\hbf(\hbf')\big)}{\hat{q}\big(\hbf(\hbf')\big)} - \alpha c\big(\hbf(\hbf'), \hbf'\big) \Big) dQ_0(\hbf') \\ 
& \geq \int \sup_{\hbf \in \Xcal} \Big( \frac{\delta_{f_{\theta}}(\hbf)}{\hat{q}(\hbf)} - \alpha c\big(\hbf, \hbf'\big) \Big) dQ_0(\hbf') \\ 
& \geq \sup_{\hat{Q}} \sup_{\gamma \in \Pi\big(\frac{P}{\hat{Q}}, \frac{P}{Q_0}\big)}  \int \Big( \frac{\delta_{f_{\theta}}(\hbf)}{\hat{q}(\hbf)} - \alpha c(\hbf, \hbf')\Big)d\gamma(\hbf, \hbf').
\end{split}
\end{equation}
Combining (\ref{eqn:claim-2-4}), (\ref{eqn:claim-2-2}) and (\ref{eqn:claim-2-1}), we see that 
\begin{equation}
\label{eqn:claim-2-5}
    \sup_{\hat{Q}} \sup_{\gamma \in \Pi\big(\frac{P}{\hat{Q}}, \frac{P}{Q_0}\big)}  \int \Big( \frac{\delta_{f_{\theta}}(\hbf)}{\hat{q}(\hbf)} - \alpha c(\hbf, \hbf')\Big)d\gamma(\hbf, \hbf') = \int \sup_{\hbf \in \Xcal}\Big(\frac{\delta_{f_{\theta}}(\hbf)}{\hat{q}(\hbf)} - \alpha c(\hbf, \hbf')\Big)dQ_0(\hbf').
\end{equation}
Finally, notice that 
\[ 
\sup_{\hat{Q}} \sup_{\gamma \in \Pi\big(\frac{P}{\hat{Q}}, \frac{P}{Q_0}\big)}  \int \Big( \frac{\delta_{f_{\theta}}(\hbf)}{\hat{q}(\hbf)} - \alpha c(\hbf, \hbf')\Big)d\gamma(\hbf, \hbf') = \sup_{\hat{Q}} \int \frac{\delta_{f_{\theta}}(\hbf)}{\hat{q}(\hbf)}d\hat{Q}(h) - \alpha W_c\big(P/\hat{Q}, P/Q_0 \big),
\]
so according to (\ref{eqn:claim-2-0}), we reach the final result:
\begin{equation}
\label{eqn:claim-2-6}
\begin{split}
    \sup_{\hat{P}:W_c(\hat{P},P^*)\leq \rho} \int \delta\big(y,f_{\theta}(\xbf, \zbf)\big) d\hat{P}(\hbf) &= \inf_{\alpha \geq 0} \Big\{\alpha \rho + \int \sup_{\hbf \in \Xcal}\Big(\frac{\delta_{f_{\theta}}(\hbf)}{\hat{q}(\hbf)} - \alpha c(\hbf, \hbf')\Big)dQ_0(\hbf')\Big\} \\
    &=\inf_{\alpha \geq 0} \Big\{ \alpha\rho + \sup_{\hat{Q}} \int \frac{\delta_{f_{\theta}}(\hbf)}{\hat{q}(\hbf)}d\hat{Q}(h) - \alpha W_c\big(P/\hat{Q}, P/Q_0 \big) \Big\}.
\end{split}
\end{equation}
\end{proof}

To reach the relaxation given in (5), we use the alternate expression for the Wasserstein distance obtained from the Kantorovich-Rubinstein duality \cite{villani2008optimal}. We denote the Lipschitz continuity for a function $f$ by $\|f\|_{L\leq l}$. When the cost function $c$ is $l$-Lipschitz continuous, $W_c(P_1,P_2)$ is also referred to as the Wasserstein-$l$ distance. Without loss of generality, we consider $\|c\|_{L\leq 1}$ such as the $\ell_2$ norm, and with that the Wasserstein distance is equivalent to:
\begin{equation}
    W_c\big(P/\hat{Q}, P/Q_0\big) = \sup_{\|f\|_{L\leq 1}}\big\{ \Ebb_{\hbf \sim P/\hat{Q}} f(\hbf) - \Ebb_{\hbf \sim P/Q_0} f(\hbf) \big\},
\end{equation}
where $f: \Xcal \to \Rbb$. In practice, when $P$ is the empirical distribution that assigns uniform weights to all the samples, we have
\begin{equation}
\begin{split}
   W_c\big(P_n/\hat{Q}, P_n/Q_0\big) &=  \sup_{\|f\|_{L\leq 1}}\big\{ \Ebb_{\hbf \sim P_n/\hat{Q}} f(\hbf) - \Ebb_{\hbf \sim P_n/Q_0} f(\hbf) \big\} \\ 
   & = \sup_{\|f\|_{L\leq 1}}\Big\{ a_1 \Ebb_{\hbf \sim P_n} \frac{f(\hbf)}{\hat{q}(\hbf)} - a_2\Ebb_{\hbf \sim P_n} \frac{f(\hbf)}{q_0(\hbf)} \Big\} \\
   & = \sup_{\|f\|_{L\leq 1}} \Ebb_{\hbf \sim P_n}\Big[\frac{f(\hbf)}{\hat{q}(\hbf)\cdot q_0(\hbf)} \big( a_1q_0(\hbf) -  a_2\hat{q}(\hbf)\big)\Big] \\
   & \leq \sup_{\hbf \in \Xcal}\Big\{\frac{1}{\hat{q}(\hbf)\cdot q_0(\hbf)}\Big\} \cdot \sup_{\|f\|_{L\leq 1}} \big\{ a_3\Ebb_{\hbf \sim P_n\cdot Q_0}f(\hbf) - a_4\Ebb_{\hbf \sim P_n\cdot \hat{Q}}f(\hbf) \big\} \\
   & \leq \frac{1}{\mu^2} \sup_{\|f\|_{L\leq \max\{a_5,a_6\}}} \big\{\Ebb_{\hbf \sim Q_0} f(\hbf) - \Ebb_{\hbf \sim \hat{Q}} f(\hbf) \big\} \\
   & = \frac{1}{\mu^2} W_{\tilde{c}}(\hat{Q}, Q_0),
\end{split}
\end{equation}
where the-above $a_i$ are all constants induced by using the change-of-measure with important-weighting estimators, and the induced cost function $\tilde{c}$ on the last line satisfies $\|\tilde{c}\|_{L\leq \max\{a_5,a_6\}}$. Therefore, we see that the Wasserstein distance between $P_n/\hat{Q}$ and $P_n/Q_0$ can be bounded by $W_{\tilde{c}}(\hat{Q}, Q_0)$. Hence, for each $\alpha \geq 0$ in (\ref{eqn:claim-2-6}), 
\[
\sup_{\hat{Q}} \Ebb_{P}\Big[ \frac{\delta(Y, f_{\theta}(\Xbf, \Zbf))}{\hat{q}(O=1|\Xbf, \Zbf)} \Big] - \tilde{\alpha} W_{\tilde{c}}(\hat{Q}, Q_0), \quad \tilde{\alpha} \geq 0,
\]
is a relaxation of the result in Claim 2. In practice, the specific forms of the cost functions $c$ or $\tilde{c}$ do not matter, because the Wasserstein distance is intractable and we use the data-dependent surrogates that we discuss in Section 3.2.

\section{Proof for Theorem 1}
\label{append:thm1}
\begin{proof}
Following the same arguments from the proof in Claim 2, we obtain the similar result stated in (\ref{eqn:claim-2-6}) that
\begin{equation}
\label{eqn:thm-1}
\begin{split}
    & \sup_{g_{\psi} \in \tilde{\Gcal}(\rho)}  \Ebb_{P} \Big[ \frac{\delta(Y, f_{\theta}(\Xbf, \Zbf))}{G\big(g_{\psi}(\Xbf, \Zbf)\big)} \Big] \\ 
    & \leq \inf_{\gamma \geq 0} \Big\{ \gamma\rho + \int \sup_{\hbf \in \Xcal}\Big(\frac{\delta_{f_{\theta}}(\hbf)}{\hat{q}(\hbf)} - \gamma c(\hbf, \hbf')\Big)  dP(\hbf) \Big\} \\
    & = \inf_{\gamma \geq 0}\Big\{ \gamma \rho + \Ebb_{P} \big[ \Delta_{\gamma}\big(f_{\theta}; \Hbf \big) \big]  \Big\}\quad (\text{by the definition of }\Delta_{\gamma}) \\
    & \leq \inf_{\gamma \geq 0}\Big\{ \gamma \rho + \Ebb_{P_n} \big[ \Delta_{\gamma}\big(f_{\theta}; \Hbf \big) \big] + \sup_{f_{\theta} \in \Fcal}\big(\Ebb_{P} \big[ \Delta_{\gamma}\big(f_{\theta}; \Hbf \big) \big] - \Ebb_{P_n} \big[ \Delta_{\gamma}\big(f_{\theta}; \Hbf \big) \big] \big) \Big\}. 
\end{split}
\end{equation}
Let $W_{\gamma} = \sup_{f_{\theta} \in \Fcal}\big(\Ebb_{P} \big[ \Delta_{\gamma}\big(f_{\theta}; \Hbf \big) \big] - \Ebb_{P_n} \big[ \Delta_{\gamma}\big(f_{\theta}; \Hbf \big) \big] \big)$, then notice that 
\[ 
W_{\gamma} = \frac{1}{n}\sup_{f_{\theta} \in \Fcal}\Big[ \sum_{i=1}^N \Ebb_{P} \big[ \Delta_{\gamma}\big(f_{\theta}; \Hbf \big)\big] - \Delta_{\gamma}\big(f_{\theta}; \Hbf_i \big) \Big] \quad \gamma \geq 0.
\]
Since $|\delta_{f_{\theta}}(\hbf)| \leq \mu M$ holds uniformly, according to the McDiarmid's inequality on bounded random variables, we first have
\begin{equation}
\label{eqn:thm-2}
p\Big( W_{\gamma} - \Ebb W_{\gamma} \geq \mu M\sqrt{\frac{\log 1/\epsilon}{2N}} \Big) \leq \epsilon.
\end{equation}
Then let $\epsilon_1,\ldots,\epsilon_N$ be the i.i.d Rademacher random variables independent of $\Hbf$, and $\Hbf'_i$ be the i.i.d copy of $\Hbf_i$ for $i=1,\ldots,N$. 

Applying the symmetrization argument, we see that
\begin{equation}
\begin{split}
    \Ebb W_{\gamma} &= \Ebb \Big[\sup_{f_{\theta}\in\Fcal} \Big| \sum_{i=1}^N \Delta_{\gamma}\big(f_{\theta}; \Hbf'_i \big)  - \sum_{i=1}^N \Delta_{\gamma}\big(f_{\theta}; \Hbf_i \big)  \Big| \Big] \\
    & = \Ebb \Big[\sup_{f_{\theta}\in\Fcal} \Big| \frac{1}{N} \sum_{i=1}^N \epsilon_i \Delta_{\gamma}\big(f_{\theta}; \Hbf'_i \big)  - \frac{1}{N} \sum_{i=1}^N \Delta_{\gamma}\big(f_{\theta}; \Hbf_i \big) \Big| \Big] \\
    & \leq 2 \Ebb\Big[ \sup_{f_{\theta}\in \Fcal} \Big|\frac{1}{N} \sum_{i=1}^N \epsilon_i \Delta_{\gamma}\big(f_{\theta}; \Hbf_i \big) \Big| \Big].
\end{split}
\end{equation}
It is clear that each $\epsilon_i \Delta_{\gamma}\big(f_{\theta}; \Hbf_i \big)$ is zero-mean, and now we show that it is sub-Gaussian as well. 

For any two $f_{\theta}, f'_{\theta}$, we show the bounded difference:
\begin{equation}
\begin{split}
    & \Ebb\Big[ \exp\Big(\lambda \big(\frac{1}{\sqrt{N}} \epsilon_i \Delta_{\gamma}\big(f_{\theta}; \Hbf_i \big) - \frac{1}{\sqrt{N}} \epsilon_i \Delta_{\gamma}\big(f'_{\theta}; \Hbf_i \big)\big)\Big)  \Big] \\
    & = \Big(\Ebb \Big[ \exp\Big( \frac{\lambda}{\sqrt{N}}\epsilon_1 \big(\Delta_{\gamma}\big(f_{\theta}; \Hbf_1 \big) - \Delta_{\gamma}\big(f'_{\theta}; \Hbf_1 \big) \big)  \Big)\Big]\Big)^N \\
    & = \Big(\Ebb \Big[ \exp\Big( \frac{\lambda}{\sqrt{N}}\epsilon_1 \big(\sup_{\hbf'}\inf_{\hbf''}\big\{ \frac{\delta_{f_{\theta}}(\hbf')}{q(\hbf')} - \gamma c(\Hbf_1,\hbf') - \frac{\delta_{f'_{\theta}}(\hbf'')}{q(\hbf'')} \big\} + \gamma c(\Hbf_1,\hbf'') \big) \Big)\Big]\Big)^N \\
    & \leq \Big(\Ebb \Big[ \exp\Big( \frac{\lambda}{\sqrt{N}}\epsilon_1 \big(\sup_{\hbf'}\big\{ \frac{\delta_{f_{\theta}}(\hbf')}{q(\hbf')} - \frac{\delta_{f'_{\theta}}(\hbf')}{q(\hbf')}   \big\}\big) \Big)\Big]\Big)^N \\
    & \leq \exp \Big(\lambda^2 \Big\|\frac{\delta_{f_{\theta}}}{q} -    \frac{\delta_{f'_{\theta}}}{q} \Big\|_{\infty}^2 \Big/2\Big) \quad (\text{by Hoeffding's inequality}).
\end{split}
\end{equation}
Hence we see that $\frac{1}{\sqrt{N}}\epsilon_i \Delta_{\gamma}\big(f_{\theta}; \Hbf_i \big)$ is sub-Gaussian with respect to $\Big\|\frac{\delta_{f_{\theta}}}{q} -    \frac{\delta_{f'_{\theta}}}{q} \Big\|_{\infty}^2$. Therefore, $\Ebb W_{\gamma}$ can be bounded by using the standard technique for Rademacher complexity and Dudley's entropy integral \cite{talagrand2014upper}:
\begin{equation}
\label{eqn:thm-3}
\Ebb W_{\gamma} \leq \frac{24}{N}\Jcal (\tilde{\Fcal}). 
\end{equation}
Combining all the above bounds in (\ref{eqn:thm-1}), (\ref{eqn:thm-2}) and (\ref{eqn:thm-3}) we obtain the desired result.
\end{proof}

\section{Proof for Corollary 1}
\label{append:corollary1}
\begin{proof}
To obtain the first result, let the data-dependent $\gamma_n$ be given by
\[ 
\gamma_n = \max_{i}\sup_{\hbf' \in \Hcal} \Big( \frac{\delta_{f_{\theta}}(\hbf')}{q(\hbf')} - \frac{\delta_{f_{\theta}}(\hbf_i)}{q(\hbf_i)}\Big) \Big/ {c(\hbf_i, \hbf')}.
\]
Then according to the definition of $\Delta_{\gamma}$, we have
\[ 
\Ebb_{P_n}\Delta_{\gamma_n}\big(f_{\theta}; \Hbf\big) = \frac{1}{N}\sum_{i}\sup_{\hbf' \in \Xcal} \Big\{\frac{\delta_{f_{\theta}}\big(\hbf' \big)}{q(\hbf')} - \max_{j}\sup_{\hbf'' \in \Xcal} \Big\{\frac{ \frac{\delta_{f_{\theta}}(\hbf'')}{q(\hbf'')} - \frac{\delta_{f_{\theta}}(\hbf_j)}{q(\hbf_j)}}{c(\hbf_j, \hbf'')} \Big\} c\big( \hbf_i, \hbf'\big)  \Big\}.
\]
It is easy to verify that
\[ 
\Ebb_{P_n}\Delta_{\gamma_n}\big(f_{\theta}; \Hbf\big) \leq \frac{1}{N}\sum_{i}\sup_{\hbf' \in \Xcal} \Big\{ \frac{\delta_{f_{\theta}}\big(\hbf' \big)}{q(\hbf')}\Big\} + \frac{\delta_{f_{\theta}}\big(\hbf_i \big)}{q(\hbf_i)} - \sup_{\hbf'' \in \Xcal} \Big\{ \frac{\delta_{f_{\theta}}\big(\hbf'' \big)}{q(\hbf'')}\Big\} = \frac{1}{N}\sum_{i} \frac{\delta_{f_{\theta}}\big(\hbf_i \big)}{q(\hbf_i)},
\]
as well as 
\[ 
\Ebb_{P_n}\Delta_{\gamma_n}\big(f_{\theta}; \Hbf\big) \geq \frac{1}{N}\sum_{i}\sup_{\hbf' \in \Xcal} \Big\{ \frac{\delta_{f_{\theta}}\big(\hbf' \big)}{q(\hbf')}\Big\} - \max_{j}\sup_{\hbf'' \in \Xcal} \Big\{\frac{ \frac{\delta_{f_{\theta}}(\hbf'')}{q(\hbf'')} - \frac{\delta_{f_{\theta}}(\hbf_j)}{q(\hbf_j)}}{c(\hbf_j, \hbf'')}c\big( \hbf_i, \hbf_j \big) \Big\},
\]
which also equals to $\frac{1}{N}\sum_{i} \frac{\delta_{f_{\theta}}\big(\hbf_i \big)}{q(\hbf_i)}$. Therefore, when $\gamma = \gamma_n$, we have $\Ebb_{P_n}\big[\Delta_{\gamma_n}\big(f_{\theta}; \Hbf\big)\big] = \Ebb_{P_n}\Big[\frac{\delta_{f_{\theta}}\big(\Hbf_i \big)}{q(\Hbf_i)}\Big]$. Similarly, it can be shown that when $\gamma > \gamma_n$, the above equality also holds. Hence, we replace $\Ebb_{P_n}\big[\Delta_{\gamma_n}\big(f_{\theta}; \Hbf\big)\big]$ with $\Ebb_{P_n}\Big[\frac{\delta_{f_{\theta}}\big(\Hbf_i \big)}{q(\Hbf_i)}\Big]$ in Theorem 1 and obtain the first result. 

To obtain the second result, we define the transportation map \cite{villani2008optimal}:
\[ 
T_{\gamma}\big(f_{\theta};\hbf \big) = \arg\max_{\hbf' \in \Xcal} \big\{ \frac{\delta_{f_{\theta}}\big(\hbf' \big)}{q(\hbf')} - \gamma c(\hbf, \hbf')\big\}.
\]
Then according to (\ref{eqn:claim-2-6}), the empirical maximizer for $\sup_{\hat{P}:W_c(\hat{P},P^*)\leq \rho} \int \delta\big(y,f_{\theta}(\xbf, \zbf)\big) d\hat{P}(\hbf)$ is attained by $\hat{P}(f_{\theta}) = \frac{1}{N}\sum_{i=1}^N I_{T_{\gamma}\big(f_{\theta};\hbf_i \big)}$ where $I_{\hbf}$ assign point mass at $\hbf$, since it maximizes $\int \sup_{\hbf \in \Xcal}\Big(\frac{\delta_{f_{\theta}}(\hbf)}{\hat{q}(\hbf)} - \gamma c(\hbf, \hbf')\Big)dQ_0(\hbf')$.

Then we let $\rho_n(f_{\theta}) = W_c(\hat{P}(f_{\theta}), P_n)$, which equals to $\Ebb_{P_n}\Big[ c\big(T_{\gamma}\big(f_{\theta};\Hbf \big), \Hbf\big) \Big]$ by definition. So now we have 
\[c_1\gamma\rho_n(f_{\theta}) + \Ebb_{P_n}[\Delta_{\gamma}\big(f_{\theta}; \Hbf \big)] = \sup_{P:W_c(P,P_n)\leq \tilde{\rho}}\Ebb_{P}\Big[\delta\big(f_{\theta}; \Hbf\big) / {q(\Hbf)}\Big],
\]
for some $\tilde{\rho}$ that absorbs the excessive constant terms. We plug it into the Theorem 1 and obtain the second result for the Corollary.
\end{proof}

\section{Implications from Tukey's Factorization on Unobserved Factors for Exposure}
\label{append:tukey}

Here, we discuss the Tukey's factorization which motivates our $G_{\beta}$ model to handle the unobserved factors in recommender system. 

We first introduce the notation of \emph{counterfactual outcome}: $Y_{u,i}(o)$, $o \in \{0,1\}$, which represents what the user feedback would be if the exposure $O_{u,i}$ were given by $o \in \{0,1\}$. In the factual world, we only get to observe $Y_{u,i}$ for either $O_{u,i}=1$ or $O_{u,i}=0$, and the tuple $\big(Y_{u,i}(1), Y_{u,i}(0)]\big)$ is never jointly observed at the same time, which to this extent connects the causal inference problem to missing data literature. 

In the absence of unobserved factor, the joint distribution of $(Y_{u,i}(1), Y_{u,i}(0))$ has a straightforward formulation and can be estimated effectively from data using tools from causal inference \cite{austin2011introduction}. However, when unobserved factor exists, there are confounding between $(Y_{u,i}(1)$ and $Y_{u,i}(0))$, which violates a fundamental assumption of many causal inference solutions.

The Tukey's factorization, on the other hand, characterizes our missing data distribution regardless of the unobserved factors as:
\begin{equation}
    p_{\beta}\big(Y(o), O | \Xbf, \Zbf \big) = p\big(Y(o) | O=o, \Xbf, \Zbf  \big) p\big(O=o | \Xbf, \Zbf \big)\cdot \frac{p_{\beta}\big(O | Y(o), \Xbf, \Zbf \big)}{p_{\beta}\big(O=o | Y(o), \Xbf, \Zbf \big)}, o\in \{0,1\},
\end{equation}
where $\frac{p_{\beta}\big(O | Y(o), \Xbf, \Zbf \big)}{p_{\beta}\big(O=o | Y(o), \Xbf, \Zbf \big)}$ concludes the unknown mechanism in the missing data distribution \cite{franks2016non,franks2019flexible}. To see how the \emph{counterfactual outcome} is reflected in the above formulation, when $O=\tilde{o} := 1-o$ and $o=1$, we have:
\[ 
p_{\beta}\big(Y(1), O=0 | \Xbf, \Zbf \big) = p\big(Y(1) | O=1, \Xbf, \Zbf  \big) p\big(O=1 | \Xbf, \Zbf \big)\cdot \frac{p_{\beta}\big(O=0 | Y(1), \Xbf, \Zbf \big)}{p_{\beta}\big(O=1 | Y(1), \Xbf, \Zbf \big)},
\]
which gives the joint distribution of the outcome if the item was not exposed and the observed data where the item is exposed.
Notice that both $p\big(Y(o) | O=o, \Xbf, \Zbf  \big)$ and $p\big(O=o | \Xbf, \Zbf \big)$ can be estimated from the data, since $Y(o)$ is observed under $O=o$. So the only unknown mechanism in the missing data distribution is: \[{p_{\beta}\big(O | Y(o), \Xbf, \Zbf \big)}\big/{p_{\beta}\big(O=o | Y(o), \Xbf, \Zbf \big)}.\]

Hence, we see the \emph{counterfactual outcome} distribution can be given by:
\begin{equation}
p_{\beta}\big(Y(o) | O=1-o, \Xbf, \Zbf  \big) \propto p_{\text{obs}}\big(Y(o)|O=o,\Xbf,\Zbf\big) \big/ G_{\beta}\big(Y(o), \Xbf, \Zbf\big), \quad o\in\{0,1\},
\end{equation}
where $p_{\text{obs}}$ denotes the observable distribution and $G_{\beta}\big(Y(o), \Xbf, \Zbf\big) = \frac{p_{\beta}\big(O=o | Y(o), \Xbf, \Zbf \big)}{p_{\beta}\big(O | Y(o), \Xbf, \Zbf \big)}$ characterizes the exposure mechanism even when unobserved factors exist. 

We treat the unknown $G_{\beta}\big(Y(o), \Xbf, \Zbf\big)$ as a learnable objective in our setting. We have discussed in Section 3.2 that we use $g_{\psi}$ to characterize the role of $\Xbf$ and $\Zbf$ in the exposure mechanism $G_{\beta}$, and hence we reach our formulation of ${\delta\big(Y, f_{\theta}(\Xbf, \Zbf)\big)}\big/{G_{\beta}\big(Y, g_{\psi} (\Xbf, \Zbf) \big)}$ in (\ref{eqn:minimax-5}). 

It has been discussed in \cite{liang2016causal} that including $Y$ in modelling the exposure mechanism may cause the so-called self-selection problem in causal inference.
Our setting does not fall into that category, since our only objective is to learn the $f_{\theta}$, rather than making inference on its treatment effect. 

We also show in the our ablation studies that if the user feedback $Y$ is not included, i.e. $G_{\beta}\big(Y, g_{\psi} (\Xbf, \Zbf) \big) := \sigma( g_{\psi} (\Xbf, \Zbf))$, the improvements over the original models will be less significant.

\section{Experiment Settings and Complete Results}
\label{append:experiment}
We provide the data descriptions, preprocessing steps, train-validation-test split, simulation settings, detailed model configuration as well as the implementation procedure in this part of the appendix. We visualize the training process that reveals the adversarial nature of our proposed approach. We then provide the a complete set of ablation study and sensitivity analysis results to demonstrate the robustness of our approach.

The implementation and datasets have been made public on GitHub\footnote{https://github.com/StatsDLMathsRecomSys/Adversarial-Counterfactual-Learning-and-Evaluation-for-Recommender-System}.

\subsection{Real-world datasets}
We consider three real-world datasets that covers movie, book and music recommendation.
\begin{itemize}
    \item \textbf{Movielens-1M }\footnote{http://files.grouplens.org/datasets/movielens/ml-1m.zip}. The benchmark dataset records users' ratings for movies, which consists of around 1 millions ratings collected from 60,40 users on 3,952 movies. The rating is from 1 to 5, and a higher rating indicates more positive feedback.
    
    \item \textbf{LastFM} \footnote{http://files.grouplens.org/datasets/hetrec2011/hetrec2011-lastfm-2k.zip}. The LastFM dataset is a benchmark dataset for music recommendation. For each of the 1,892 listeners, they tag the artists they may find fond of over time. Since the tag is a binary indicator, the LastFM is an implicit feedback dataset.  There is a total of 186,479 tagging events, where 12,523 artists have been tagged. 
    
    \item \textbf{GoodReads} \footnote{https://sites.google.com/eng.ucsd.edu/ucsdbookgraph/home}. The benchmark book recommendation dataset is scraped from the users' public shelves on \emph{Goodread.com}. We use the user review data on the \emph{history} and \emph{biography} sections due to their richness. There are in total 238,450 users, 302,346 unique books, and 2,066,193 ratings in these sections. The rating range is is also from 1 to 5, a higher rating indicates more positive feedback.
\end{itemize}

\subsection{Data preprocessing and train-validation-test splitting}

The Movielens-1M dataset has been filtered before made public, where each user in the dataset has rated at least 20 movies. For the LastFM and Goodread datasets, we first eliminate infrequent items (books/artists) and users that have less than 20 records. After examination, we find a small proportion of users having an abnormal amount of interactions. Therefore, we treat the users who have more than 1,000 interactions as spam users and not include them into our analysis.

The train-validation-test split is carried out based on the order of the user-item interactions. We adopt the standard setting, where for each user interaction sequence, all items but the last two are used in training, the second-to-last interaction is used in validation, and the last interaction is used in testing.

\subsection{Simulation settings}
\label{append:simulation}
In a modern real-world recommender system, the exposure mechanism is determined by the underlying recommender model as well as various other factors. In an attempt to mimic the real-world recommender systems, we design a \emph{two-stage} simulation approach to generate the semi-synthetic data that remains truthful to the signal in the original dataset. 

The purpose of the first stage is to learn the characteristic from the data, such as the user relevance (rating) model and the partial exposure model (which may be inaccurate due to the partial-observation of exposure status). In the second stage, we simulate the working method of a real-world recommender system and generate the user response accordingly. Since we wish to recover the user-item relevance as accurate as possible, we choose to use the explicit feedback dataset for our simulation, i.e. the \emph{Movielens-1M} and \emph{Goodreads} dataset.

In the first stage, given a true rating matrix, we train two hidden-factor matrix factorization models. The first model tries to recover the rating matrix and by minimizing the mean-squared loss. We refer to this model as the \emph{relevance model}. Since for the explicit feedback data, the rated items must have all been exposed, so given the output $\hat{\Ebb}[R_{u,i}|O_{u,i}=1]$, we define the relevance probability as \[p_{\text{sim1}}(Y_{u,i}=1|O_{u,i}=1) := \sigma \big(\hat{\Ebb}[R_{u,i}|O_{u,i}=1] + \epsilon_1 \big),\] where $\sigma(\cdot)$ is the sigmoid function and the Gaussian noise $\epsilon_1$ reflects the perturbations brought by unobserved factors. The second model is an implicit-feedback model trained to predict the occurrence of the rating event $\hat{p}(O_{u,i}=1)$, where instead of using the original ratings, the non-zero entries in the rating matrix are all converted to one. 

After obtaining the $\hat{p}(O_{u,i}=1)$, we define the simulation exposure probability as $\log p_{\text{sim1}}(O_{u,i}=1) = \log \hat{p}(O_{u,i}=1) + \epsilon_2$, where $\epsilon_2$ also gives the extra randomness due to the unobserved factors. 

Now, after obtaining the simulated $p_{\text{sim}}(Y_{u,i}=1|O_{u,i}=1)$ and $p_{\text{sim}}(O_{u,i}=1)$, which reflects both the relevance and exposure underlies the real data generating mechanism while taking account of the effects from unobserved factors, we generate the first-stage click data based by: 
\[p_{\text{sim1}}(Y_{u,i}=1) = p_{\text{sim1}}(Y_{u,i}=1|O_{u,i}=1) p_{\text{sim1}}(O_{u,i}=1).\]

So far, in the first stage, we have generated an implicit feedback dataset that remains truthful to the original real dataset. Now we add the self-defined components that gives us more control over the exposure mechanism. Specifically, we obtain the new user and item hidden factors $\xbf, \zbf$ by training another implicit matrix factorization model using the generated click data. We generate the extra self-defined exposure function $e(\xbf, \zbf)$, and add it to the first-stage $p_{\text{sim1}}$ and obtain the second-stage exposure mechanism:
\[\log p_{\text{sim2}}(O_{u,i}=1) = \log p_{\text{sim1}}(O_{u,i}=1) + e(\xbf, \zbf).\]
The final click data is then generated via: \[p_{\text{sim2}}(Y_{u,i}=1) = p_{\text{sim1}}(Y_{u,i}=1|O_{u,i}=1) p_{\text{sim2}}(O_{u,i}=1).\]

We point out that having the second stage in the simulation is important, because the focus of the first stage is to mimic the generating mechanism of the real-world dataset. The second stage allows us to control the exposure mechanism via the extra $e(\xbf, \zbf)$. Also, retraining the implicit matrix factorization model in the beginning of the second stage is not required, thought it helps us to better characterize the data generated in the first stage.

\subsection{Model configuration and implementation}

For all the baseline models we consider here (other than \textbf{Pop}), the dimension of the user and item hidden factors, initial learning rate and the $\ell_2$ regularization strength are the basic hyperparameters. We select the initial learning rate from \{0.001, 0.005, 0.01, 0.05, 0.1\}, and the $\ell_2$ regularization strength from \{0, 0.01, 0.05, 0.1, 0.2, 0.3\}. The tuning parameters are selected separately to avoid excessive computations. We fix the hidden dimension at 32 for our models in order to achieve fair comparisons in the experiments. Also, notice that our approach has approximately twice the number of parameters with respect to the corresponding baseline model. In practice, the hidden dimension can be treated as a hyperparameter as well. We provide sensitivity analysis on the hidden dimension later in this section. We use the $Hit@10$ on validation data as the metric for selecting hyperparameters.

To make sure that the superior performance of our approach is not a consequence of higher model complexity, we double the hidden factor dimension of the baseline models to 64 when necessary.

Among the baseline models, the \textbf{Pop}, \textbf{CF} \cite{schafer2007collaborative}, \textbf{GMF} and \textbf{Neural CF} \cite{he2017neural} are all standard approaches in recommender system who have relatively simpler structures, so we adopt the default settings and do not discuss their details. We focus more on the attention-based sequential recommendation model \textbf{Attn} and the propensity-score method \textbf{PS}. For the \textbf{Attn}, we adopt the model setting from \cite{xu2019self,kang2018self} where the self-attention mechanism is added on top of a item embedding layer. We treat the hidden dimension of the key, query and value matrices, and the number of dot-product attention heads as the additional tuning parameters. For the \textbf{PS} method, there are two stages:
\begin{itemize}
    \item Obtain $g_{\psi}^*$ by minimizing $\Ebb_{P_n} \big[\delta\big(Y, g_{\psi}(\Xbf, \Zbf) \big) \big]$ as a standard ERM;
    \item Implement: $\underset{f_{\theta} \in \mathcal{F}, \beta}{\text{minimize}} \, \Ebb_{P_n}\Big[ \frac{\delta(Y, f_{\theta}(\Xbf, \Zbf))}{G_{\beta}\big(g_{\psi}^*(\Xbf, \Zbf), Y \big)} \Big]$, as a propensity-weighted ERM.
\end{itemize}
The tuning parameters for $g_{\psi}$ and $f_{\theta}$ are selected in each stage separately. 

The configurations for the proposed approach consists of two parts: the usual model configuration for $f_{\theta}$ and $g_{\psi}$, and the two-timescale train schema. Firstly, we find out that the tuning parameters selected for $f_{\theta}$ and $g_{\psi}$ when being trained alone also gives the near-optimal performance in our adversarial counterfactual training setting. Therefore, we directly adopt the hyperparameters (other than the learning rate) selected in their individual training for $f_{\theta}$ and $g_{\psi}$. We experiment on several settings for the two-timescale update. Specifically, we wish to understand the impact of the relative magnitude of the initial learning rates $r_{\theta}$ and $r_{\psi}$. In practice, we care less about the learning rate discount when using the Adam optimizer, since the learning rate is automatically adjusted. Intuitively speaking, the smaller the $r_{\psi}$ (relative to $r_{\theta}$ ), the less $g_{\psi}$ is subject to the regularization in the beginning stage, and its adversarial behavior is less restricted. As a consequence, $f_{\theta}$ may not learn anything useful. 
We provide empirical evidence to support the-above point in Figure \ref{fig:adv-1}, with the detailed discussion shown later.
Finally, the regularization parameter $\alpha$ for the proposed approach is selected from \{0.1, 1, 2\}. 

In conclusion, the hyperparameters that are specific to the proposed adversarial counterfactual training are the initial learning rates $r_{\theta}$ and $r_{\psi}$, as well as the regularization parameter $\alpha$.

\subsection{Computation}

All the models, including the matrix factorization models, are implemented with \emph{PyTorch} on a Nvidia V100 GPU machine. We use the \emph{sparse Adam}\footnote{https://agi.io/2019/02/28/optimization-using-adam-on-sparse-tensors/} optimizer to update the hidden factors, and the usual Adam optimizer to update the remaining parameters. We use sparse Adam for the hidden factors because both the user and item factor are relatively sparse in recommendation datasets. The Adam algorithm leverages the momentum of the gradients from the previous training batch, which may not be accurate for the item and user factors in the current training batch. The sparse Adam optimizer is designed to solve the above issue for sparse tensors. 

We use the early-stopping training method both for the baseline models, where we terminate the training process when the validation metric stops improving for 10 consecutive epochs. And for our approach, we monitor the minimax objective value and terminate the training process if it stops changing for more than $\epsilon=0.001$ after ten consecutive epochs. 

It is straightforward to tell that in a single update step, the space and time complexity of our proposed adversarial counterfactual training is exactly the summation for that of $f_{\theta}$ and $g_{\psi}$ (where the complexity induced by $G_{\beta}$ is almost negligible). In general, our approach may take more training epochs to converge depending on the $r_{{\theta}}/r_{\psi}$ in our two-timescale training schema.

\subsection{Visualization of the adversarial training process}

\begin{figure}[htb]
    \centering
    \includegraphics[width=0.9\textwidth]{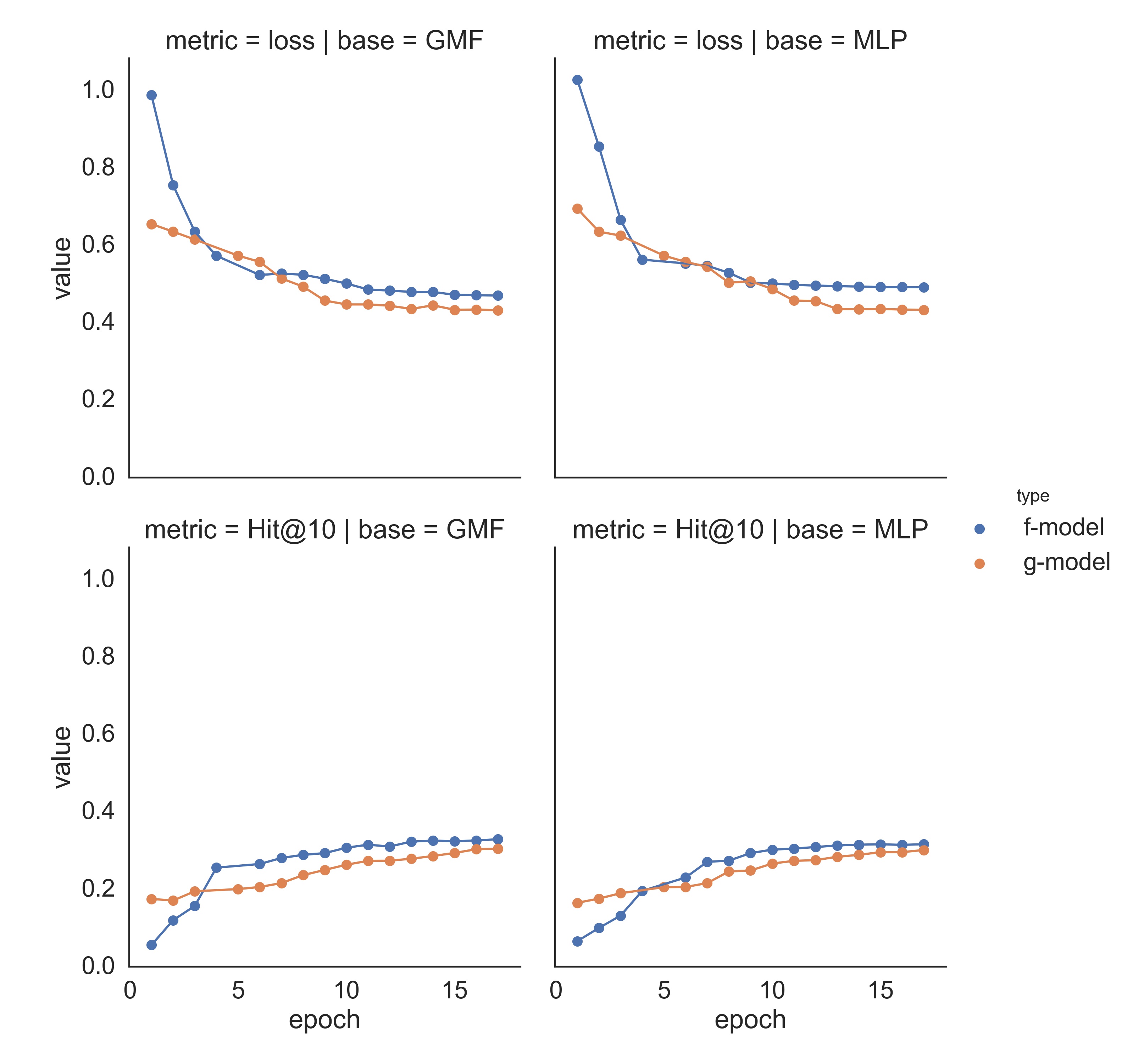}
    \caption{Adversarial training processes on the \emph{Goodread} synthetic data using ACL (GMF / GMF) and ACL (MLP / MLP) as respectively. The upper panel gives the training objective for $f_{\theta}$ and $g_{\psi}$, i.e. $\Ebb_{P_n}\Big[{\delta\big(Y, f_{\theta}(\Xbf, \Zbf) \big)}\big/{G_{\beta}(Y, g_{\psi}(\Xbf, \Zbf))} \Big]$ and $\Ebb_{P_n}\big[\delta(Y, g_{\psi}(\Xbf, \Zbf)) \big]$. The lower panel gives the evaluation metric on the validation dataset.}.
    \label{fig:adv-1}
\end{figure}

\begin{figure}[hbt]
    \centering
    \includegraphics[width=0.95\textwidth]{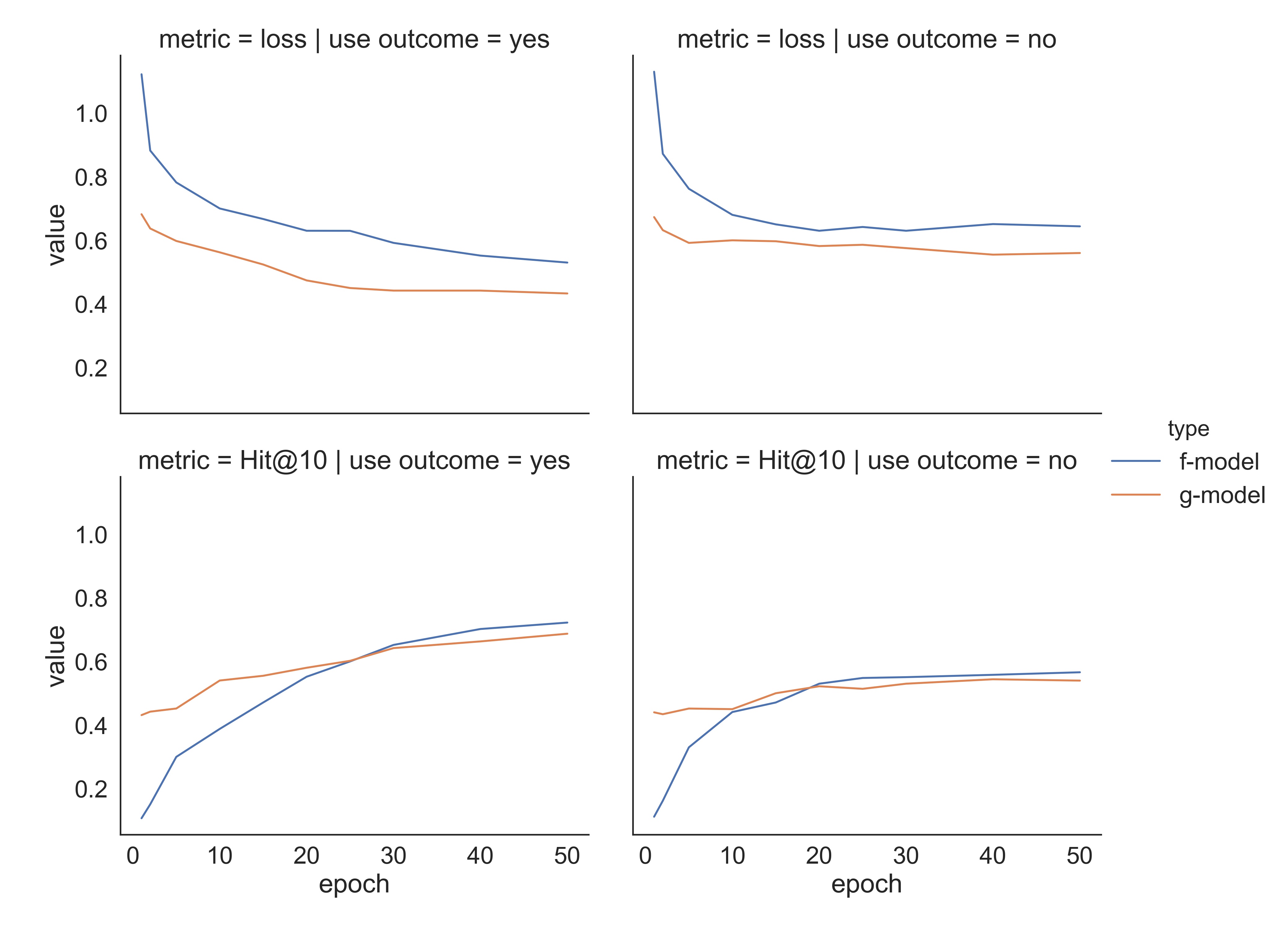}
    \caption{The adversarial training process on the real \emph{Goodread} data using ACL (Attn / Attn) shows same pattern for the sequential recommendation setting, and demonstrates the effectiveness of including the outcome into the $G_{\beta}$ for modelling the exposure mechanism. The "use outcome" indicates whether $Y$ is used for modelling $G_{\beta}$.}
    \label{fig:adv-2}
\end{figure}

To demonstrate the underlying adversarial training process of the proposed adversarial counterfactual training method, we plot the training progress under several settings in Figure \ref{fig:adv-1} and \ref{fig:adv-2}. From Figure \ref{fig:adv-1}, we observe the following things.
\begin{itemize}
    \item With a larger initial learning rate, $g_{\psi}$ tends to fit the data quicker than $f_{\theta}$.
    \item In the beginning stage, when $g_{\psi}$ has not yet fitted the data well, its adversarial behavior on $f_{\theta}$ is too strong, since both the loss value and the evaluation metric for $f_{\theta}$ is poor during that period. This also suggests the importance of using a larger initial learning rate for $g_{\theta}$.
    \item As the training progresses, $f_{\theta}$ eventually catches up with and outperforms $g_{\psi}$ in terms on the evaluation metric. However, the loss objective for $f_{\theta}$ is still larger, which is reasonable since it has the extra adversarial term in $\Ebb_{P_n}\Big[{\delta\big(Y, f_{\theta}(\Xbf, \Zbf) \big)}\big/{G_{\beta}(Y, g_{\psi}(\Xbf, \Zbf))} \Big]$, which is controlled by $g_{\psi}$. This also implies that $g_{\psi}$ is acting adversarially throughout the whole process, which matches our design of the adversarial game.
    \item The training process gradually achieves the local minimax optimal, where both $f_{\theta}$ and $g_{\psi}$ are unable to undermine the performance of each other, and their individual performances improve at the same pace in the latter training phase.
\end{itemize}

We then examine the adversarial training on the real-world dataset using the sequential recommendation model ACL (Attn / Attn). In Figure \ref{fig:adv-2}, firstly, we observe the same pattern as that of Figure \ref{fig:adv-1}, which suggests that the above discussions also apply to the real-world data and the sequential recommendation setting. 

Further, we conduct a set of experiment where the outcome is not included in modelling the exposure mechanism $G_{\beta}$. First of all, we see that the same adversarial training patterns still hold whether or not we include the outcome in modelling $G_{\beta}$. Secondly, the performances, both in terms of the loss value and evaluation metric, are less ideal when $Y$ is not included in $G_{\beta}$. 

\subsection{Complete ablation study}

Due to the space limitation, we only provided part of the ablation study in the main paper, and leave the rest to this part of the appendix. Firstly, we provide the complete results on using the propensity score model in Table \ref{tab:real-full-ps} for the three real-world datasets. 

Comparing with the results in Table 2, we see that our adversarial counterfactual training approach still outperforms their propensity score counterparts, which again emphasizes the importance of having the adversarial process between $f_{\theta}$ and $g_{\psi}$.
Secondly, we provide the full set of results for the baseline models trained with our adversarial counterfactual approach on the real-world dataset (Figure \ref{tab:real-acl}). As we mentioned in Section 5, models trained with our approach uniformly outperforms their counterparts. Notice that the superior performances of our approach do not benefit from a larger model complexity, since we have doubled the hidden factor dimension of the corresponding baseline models such that the number of parameters are approximately the same for all models.

\begin{table}[!]
    \centering
    \resizebox{1.05\columnwidth}{!}{%
    \begin{tabular}{c c c c c c c c c}
        \hline \hline
      & MLP & MLP & GMF & GMF & NCF & NCF & Attention & Attention \\
      \hdashline
       \textbf{config}  & Pop & MLP & Pop & GMF & Pop & NCF & Pop & Attention \\
      \hline
        \multicolumn{9}{c}{\textbf{\textit{MovieLens-1M}}} \\
        Hit@10 &  61.93 (.2) & 60.85 (.1) & 64.21 (.3) & 62.19 (.1) & 63.78 (.4) & 61.28 (.2) & {81.97} (.1) & 81.05 (.2)  \\
        NDCG@10 & 33.37 (.1) & 31.90 (.2) & 34.96 (.1) & 32.53 (.2) & 34.05 (.1)& 30.98 (.3) &{54.51} (.1) & 52.33 (.1) \\
        \hline
        \multicolumn{9}{c}{\textbf{\textit{Last-FM}}} \\
        Hit@10 &  82.06 (.3) & 81.32 (.1) & {82.64} (.3) & 81.87 (.1) & 82.29 (.3) & 80.35 (.2) & 72.71 (.2) & 70.98 (.1)  \\
        NDCG@10 & 57.55 (.2) & 58.16 (.1) & {58.83} (.2) & 57.92 (.3) & 58.40 (.1)& 57.02 (.3) & 60.13 (.2) & 59.33 (.2) \\
        \hline
        \multicolumn{9}{c}{\textbf{\textit{Goodreads}}} \\
        Hit@10 & 62.59 (.1)& 60.03 (.3) & 64.92 (.2) & 64.43 (.2) & 63.75 (.2) & 61.44 (.3) & {73.39} (.3) & 71.37 (.2) \\
        NDCG@10 & 38.01 (.2) & 37.32 (.1) & 39.21 (.1) & 38.45 (.1) & 38.85 (.2) & 38.03 (.1) & {49.99} (.1) & 49.18 (.2) \\
        \hline \hline
        
    \end{tabular}
    }
    \caption{\footnotesize Standard evaluations on the real-world data using the propensity-score models.}
    \label{tab:real-full-ps}
\end{table}

\begin{table}[!]
    \centering
    \resizebox{1.05\columnwidth}{!}{%
    \begin{tabular}{c c c c c c c c c}
        \hline \hline
     \textbf{ACL variant} & \multicolumn{2}{c}{ACL-MLP} & \multicolumn{2}{c}{ACL-GMF} & \multicolumn{2}{c}{ACL-NCF} & \multicolumn{2}{c}{ACL-Attention} \\
      \hdashline
       \textbf{Metric} & Hit@10 & NDCG@10 &  Hit@10 & NDCG@10 &  Hit@10 & NDCG@10 &  Hit@10 & NDCG@10 \\
      \hline
      \emph{MovieLens-1M} & 62.04 (.2) & 33.59 (.2) & 64.32 (.2) & 33.70 (.1) & 63.97 (.2) & 34.81 (.1) & 83.64 (.1) & 55.71 (.2)
      \\
      \emph{Last-FM} & 82.88 (.2) & 57.43 (.2) & 83.64 (.2) & 59.11 (.1) & 83.09 (.2) & 58.93 (.2) & 72.02 (.2) & 59.45 (.1)
      \\
      \emph{Goodreads} & 62.90 (.2) &  38.57 (.1) & 64.57 (.2) & 39.54 (.1) & 63.95 (.2) & 38.72 (.1) & 73.82 (.3) & 49.99 (.1) \\
      \hline
        
    \end{tabular}
    }
    \caption{\footnotesize Standard evaluations on the real-world data considering all ACL base model.}
    \label{tab:real-acl}
\end{table}

\subsection{Sensitivity analysis}
\begin{figure}
    \centering
    \includegraphics[width=0.9\textwidth]{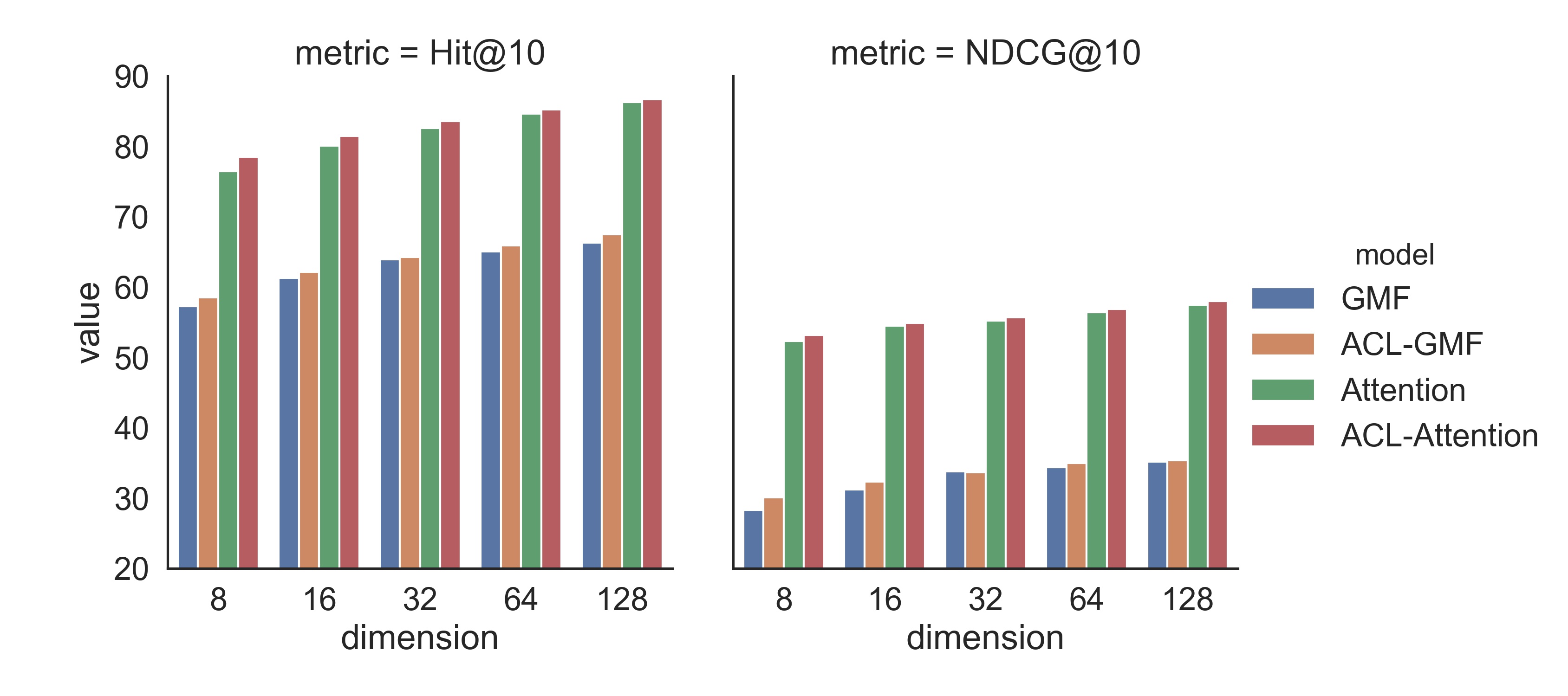}
    \includegraphics[width=0.9\textwidth]{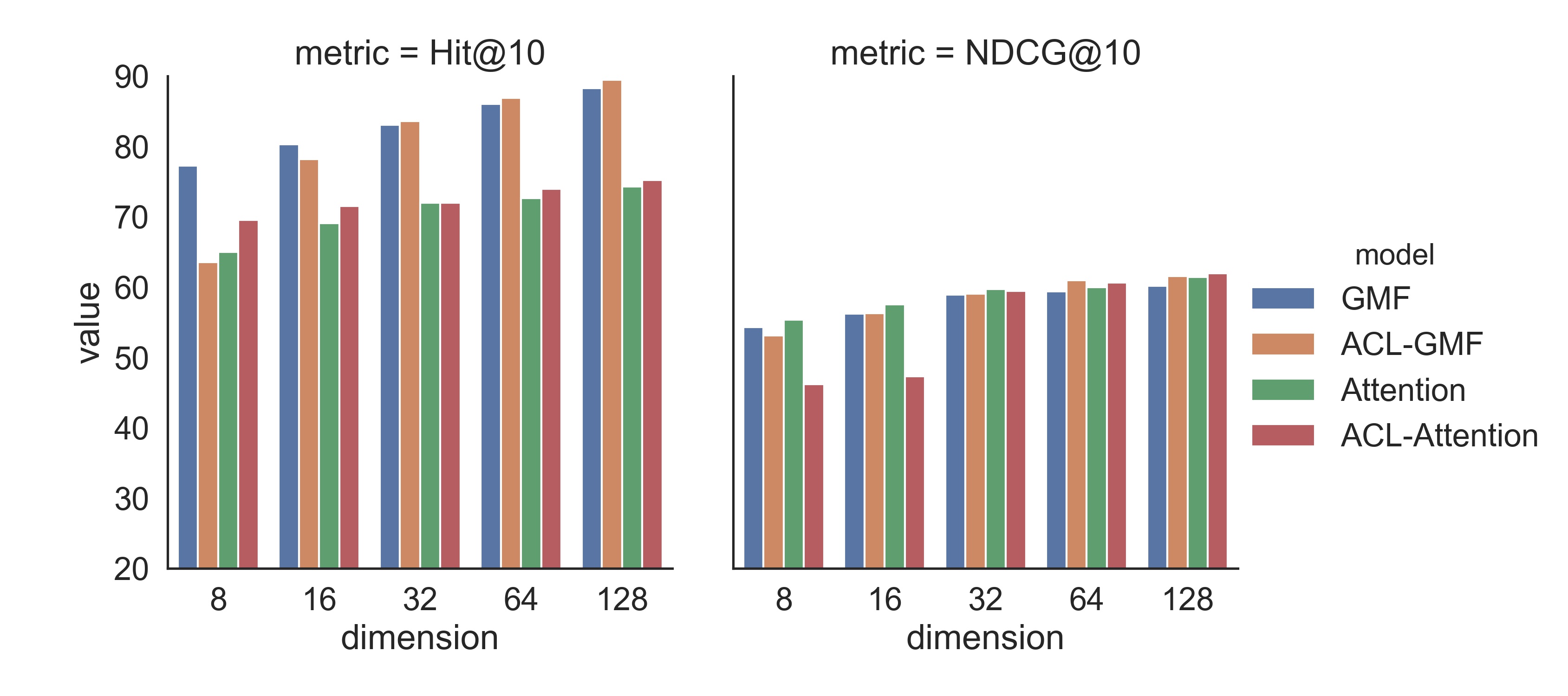}
    \includegraphics[width=0.9\textwidth]{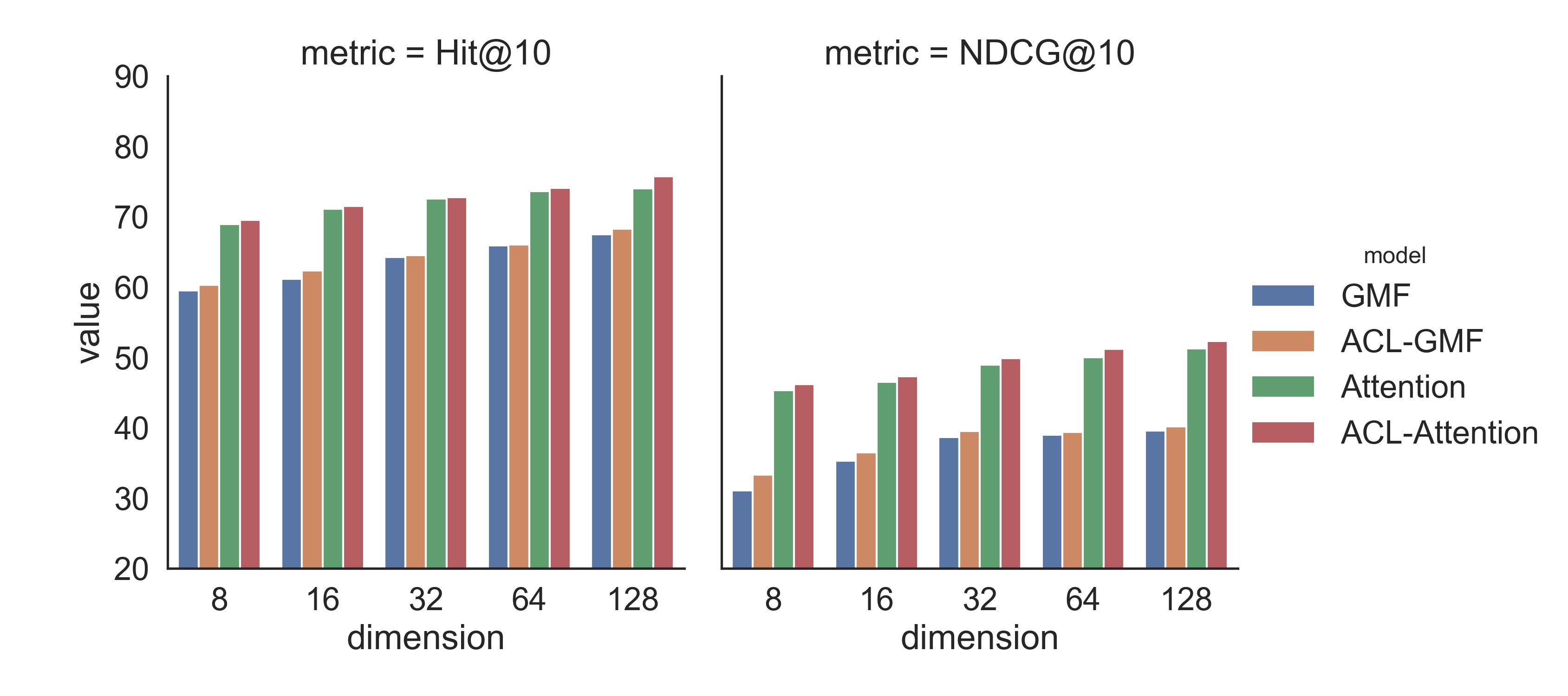}    
    \caption{Sensitivity analysis of hidden factor dimension for the content-based ACL(GMF / GMF) model and the sequential ACL(Attn / Attn) model together with their corresponding baseline models, on the three real-world datasets. Recall that the hidden dimensions for the corresponding baselines are doubled from what is shown in the plots to achieve fair comparisons. From the top to bottom are results for the \emph{Movielens-1M}, \emph{LastFM} and \emph{Goodread.com} data}
    \label{fig:dim_tune}
\end{figure}

We provide the sensitivity analysis for the proposed adversarial counterfactual approach, mostly focus on the user/item hidden factor dimension size and the regularization parameter $\alpha$. We show the results of on the real-world datasets. The sensitivity analysis on user/item hidden factor dimension size is shown in Figure \ref{fig:dim_tune}, and we observe that the larger dimensions most often lead to better outcome (within the range we consider), which is in accordance with the common consensus in the recommender system domain. This also suggests that our approach inherits some of the properties from the $f_{\theta}$ and $g_{\psi}$, so the model understanding diagnostics also become easier if $f_{\theta}$ and $g_{\psi}$ are well-studied.

\begin{figure}
    \centering
    \includegraphics[width=\textwidth]{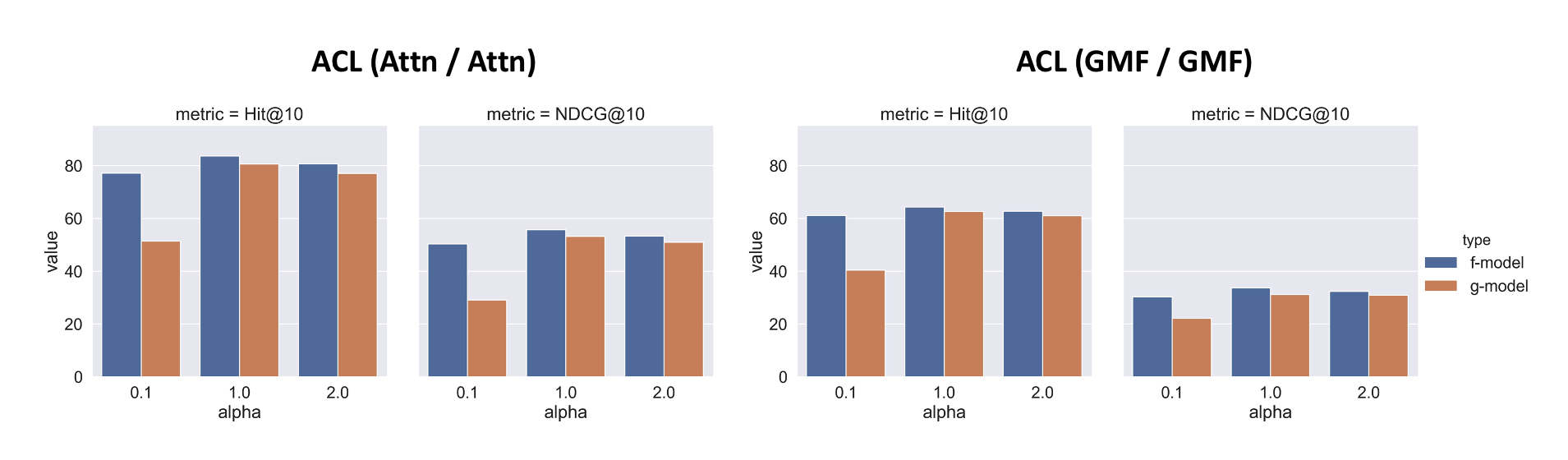}
    \includegraphics[width=\textwidth]{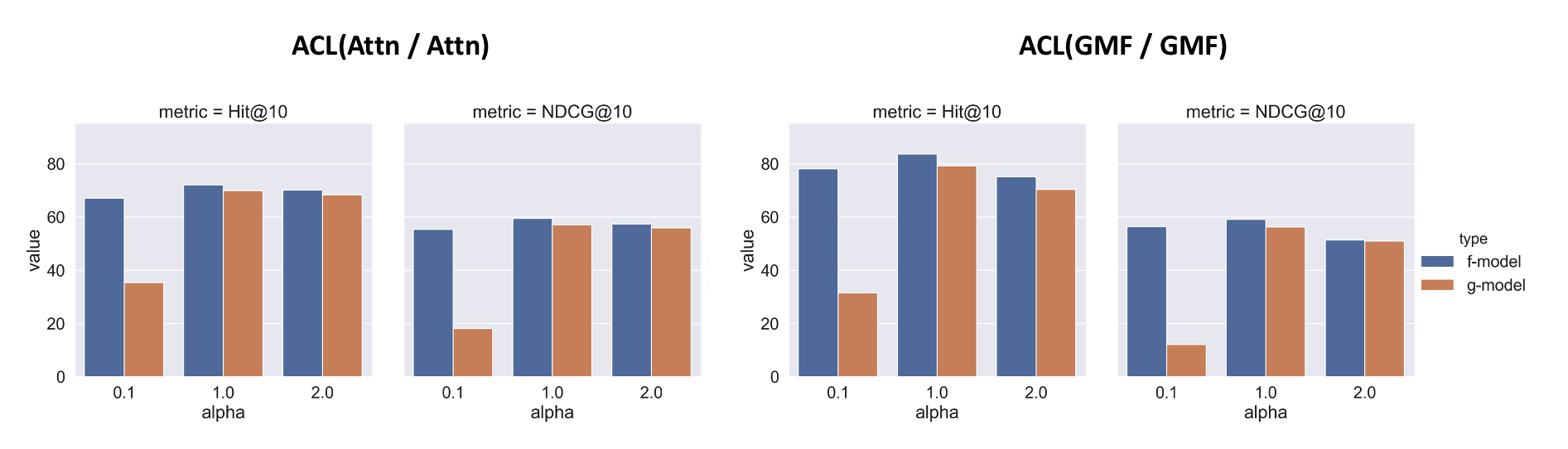}
    \includegraphics[width=\textwidth]{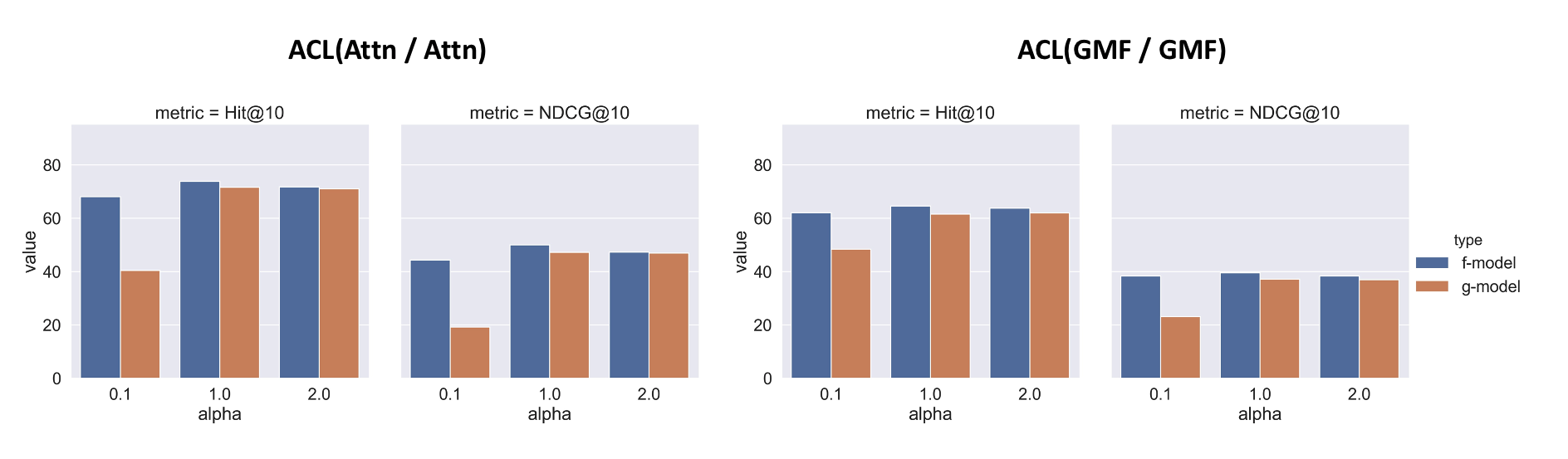}

    \caption{Sensitivity analysis on the regularization parameter  $\alpha$ for the content-based ACL (GMF / GMF) model and the sequential ACL(Attn / Attn) model for their $f_{\theta}$ and $g_{\psi}$ components, on the three real-world datasets (from the top to bottom are results for the \emph{Movielens-1M}, \emph{LastFM} and \emph{Goodread.com} data).}
    \label{fig:lambda_tune}
\end{figure}

The sensitivity analysis on the regularization parameter $\alpha$ is provided in Figure \ref{fig:lambda_tune}. We do not experiment on a wide range of $\alpha$; however, the results we have at hand already tells the patterns, that our approach achieves the best performances when $\alpha$ is neither too big nor too small. As a matter of fact, this phenomenon on regularization parameters is widely acknowledged in the machine learning community. In terms of our context, when $\alpha$ is too small, the regularization on $g_{\psi}$ becomes relatively weak compared with the loss objective of $f_{\theta}$, so $g_{\psi}$ does not fit the data well. As a consequence, $f_{\theta}$ also suffers from the under-fitting issues of $g_{\psi}$. On the other hand, when $\alpha$ gets too large, the minimax game will focus more on fitting $g_{\psi}$ to the data and overlooks $f_{\theta}$.

\subsection{Online experiment settings}
The online experiments provide valuable evaluation results that reveal the appeal of our approach for real-world applications. All the online experiments were conducted for a content-based item page recommendation module, under the implicit feedback setting where the users click or not click the recommendations. A list of ten items is shown to the customer on each item page, e.g. items that are similar or complementary to the anchor item on that page. The recommendation is personalized, so the user id and user features are included in the model as well. 

In each iteration of model deployment, the new item features and user features are added into the previous model. The major architecture of the recommendation model remains unchanged during the iterations, which makes it favorable for examining our approach. By the time we write this paper, there have been four online experiments (A/B testing) conducted for a total of eight models that are trained offline using our proposed adversarial counterfactual training, and then evaluated using the history implicit feedback data. Unobserved factors such as the real-time user features, page layout and same-page advertisements are continually changing and are thus not included in the analysis. The metric that we used to compare the different offline evaluation methods with online evaluation is the click-through rate.

\end{document}